\documentclass[prb,twocolumn,amsmath,amssymb,showpacs,longbibliography]{revtex4-1}

\usepackage{hyperref}
\pdfoutput=1
\usepackage{graphicx}
\usepackage{epsfig}
\usepackage{epstopdf}

\usepackage{amsmath}
\usepackage{bm}
\usepackage{amssymb}
\usepackage{textcomp}
\usepackage{ulem}

\usepackage{import}
\usepackage{color}
\usepackage{verbatim}
\usepackage{ulem}

\usepackage{float}

\newcommand\abs[1]{\left| #1 \right|}
\DeclareMathOperator{\sgn}{sgn}
\def\Xint#1{\mathchoice
   {\XXint\displaystyle\textstyle{#1}}%
   {\XXint\textstyle\scriptstyle{#1}}%
   {\XXint\scriptstyle\scriptscriptstyle{#1}}%
   {\XXint\scriptscriptstyle\scriptscriptstyle{#1}}%
   \!\int}
\def\XXint#1#2#3{{\setbox0=\hbox{$#1{#2#3}{\int}$}
     \vcenter{\hbox{$#2#3$}}\kern-.5\wd0}}

\def\dashint{\Xint-}

\begin{document}

\title{Raman Spectroscopy of Electrochemically-Gated Graphene Transistors: Geometrical Capacitance, Electron-Phonon, Electron-Electron, and Electron-Defect Scattering}

\author{Guillaume Froehlicher}
\affiliation{Institut de Physique et Chimie des Mat\'eriaux de Strasbourg and NIE, UMR 7504, Universit\'e de Strasbourg and CNRS, 23 rue du L\oe{}ss, BP43, 67034 Strasbourg Cedex 2, France}

\author{St\'ephane Berciaud}
\email{stephane.berciaud@ipcms.unistra.fr}
\affiliation{Institut de Physique et Chimie des Mat\'eriaux de Strasbourg and NIE, UMR 7504, Universit\'e de Strasbourg and CNRS, 23 rue du L\oe{}ss, BP43, 67034 Strasbourg Cedex 2, France}

\begin{abstract}
We report a comprehensive micro-Raman scattering study of electrochemically-gated graphene field-effect transistors. The geometrical capacitance of the electrochemical top-gates is accurately determined from dual-gated Raman measurements, allowing a quantitative analysis of the frequency, linewidth and integrated intensity of the main Raman features of graphene. The anomalous behavior observed for the G-mode phonon is in very good agreement with theoretical predictions and provides a measurement of the electron-phonon coupling constant for zone-center ($\Gamma$ point) optical phonons. In addition, the decrease of the integrated intensity of the 2D-mode feature with increasing doping, makes it possible to determine the electron-phonon coupling constant for near zone-edge (K and K' points) optical phonons. We find that the electron-phonon coupling strength at $\Gamma$ is five times weaker than at K (K'), in very good agreement with a direct measurement of the ratio of the integrated intensities of the resonant intra- (2D') and inter-valley (2D) Raman features. We also show that electrochemical reactions, occurring at large gate biases, can be harnessed to efficiently create defects in graphene, with concentrations up to approximately $1.4\times 10^{12}~\rm cm^{-2}$. At such defect concentrations, we estimate that the electron-defect scattering rate remains much smaller than the electron-phonon scattering rate.  The evolution of the G- and 2D-mode features upon doping remain unaffected by the presence of defects and the doping dependence of the D mode closely follows that of its two-phonon (2D mode) overtone. Finally, the linewidth and frequency of the G-mode phonon as well as the frequencies of the G- and 2D-mode phonons in doped graphene follow sample-independent correlations that can be utilized for accurate estimations of the charge carrier density.

\end{abstract}


\maketitle

\section{Introduction}
\label{sec1}

Graphene, as an atomically thin two-dimensional crystal, features an electron gas that is directly exposed to its local environment. As a result, graphene is uniquely sensitive to external stimuli. This is remarkably illustrated by the electric field effect, which makes it possible to swiftly tune the carrier density of graphene (\textit{i.e.}, its Fermi energy  $E_{\rm F}$) and, in return, to control a wealth of fundamental properties, among which, the electrical~\cite{Novoselov2004,Zhang2005,Novoselov2005} and optical conductivities,\cite{Wang2008,Li2008,Mak2014} as well as of the electron-phonon coupling.\cite{Yan2007,Pisana2007} From a more applied standpoint, the unique controllability of graphene can be harnessed in a variety of nano-devices.\cite{Novoselov2012}

Among the various experimental techniques employed to study graphene, Raman scattering spectroscopy~\cite{Malard2009,Ferrari2013}  stands out as a fast, sensitive, and minimally invasive tool in order to probe electron-phonon,\cite{Yan2007,Pisana2007,Berciaud2010,Chae2010,Freitag2010} electron-electron~\cite{Basko2009} and electron-defect scattering~\cite{Bruna2014,Liu2013} at variable carrier density. Raman spectroscopy is also routinely employed to characterize unintentional doping in graphene~\cite{Casiraghi2007,Berciaud2009,Ni2009} and to study the sensitivity of graphene to atmospheric~\cite{Ryu2010} and chemical dopants.\cite{Jung2009,Zhao2010,Jung2011,Howard2011,Crowther2012,Parret2013,Chen2014} Quantitative investigations of doped graphene are particularly relevant, since several interesting phenomena, such as superconductivity,\cite{McChesney2010,Profeta2012,Nandkishore2012} ferromagnetism,\cite{Ma2010} charge or spin density waves,\cite{Li2010,Makogon2011} as well as changes in the plasmon spectrum \cite{Koppens2011,Garcia2014} are expected to occur in the strong doping regime $(\abs{E_{\rm F}}\gtrsim 1~\textrm{eV})$.

In practice, solid state graphene field-effect transistors (FETs), typically using a Si substrate as a back-gate and a SiO$_2$ epilayer as a gate dielectric, have been widely used to study the Raman response of graphene in the vicinity of the Dirac point $\left(\abs{E_{\rm F}}= 0-300~\textrm{meV}\right)$.\cite{Yan2007,Pisana2007,Yan2008,Araujo2012} To access higher doping levels, other methodologies based on chemical doping~\cite{Jung2009,Zhao2010,Jung2011,Howard2011,Crowther2012,Parret2013,Chen2014} and electrochemical gating~\cite{Lu2004,Kim2013,Das2008} have been introduced. The former is highly efficient, resulting in charge carrier concentrations exceeding $10^{14}~\textrm{cm}^{-2}$, but is irreversible and little controllable. The latter, which relies on the formation of nanometer-thin electrical double layers (EDL) with high geometrical capacitance, makes it possible to reversibly attain electron or hole concentrations as high as $\sim 10^{14}~\rm cm^{-2}$, $\left(\textit{i.e.,} \abs{E_{\rm F}}\sim 1~\textrm{eV}\right)$ at cryogenic temperatures.\cite{Efetov2010} Recently, electrochemically-gated graphene FETs have been successfully employed to investigate electron-phonon coupling,\cite{Das2008,Yan2009,Das2009,Kalbac2010,Chen2011,Chattrakun2013} but also bandgap formation in bilayer graphene,\cite{Mak2009,Zhang2009} electron transport at high carrier density,\cite{Efetov2010,Efetov2011,Ye2011} many-body phenomena,\cite{Mak2014} as well as to electrically control the interaction between nano-emitters and graphene.\cite{Lee2014,Tielrooij2015}

In such studies, an accurate determination of $E_{\rm F}$ (hence of the gate capacitance) as a function of the gate voltage is a critical requirement. However, as opposed to solid state FETs, in which the oxide dielectric constant and thickness can be known with accuracy, the thickness of the electrical double layer may be highly sensitive to the device geometry, fluctuate spatially and vary over time. This further highlights the need for (i) robust methods for device fabrication and, (ii) accurate tools to experimentally measure the gate capacitance. In previous works, the gate capacitance of electrochemically gated graphene FETs has been evaluated from an estimation of the thickness of the EDL,\cite{Das2008,Das2009} from capacitance \cite{Xia2009,Uesugi2013} or Hall measurements,\cite{Ye2011,Efetov2010,Efetov2011,Bruna2014} or from optical absorption spectroscopy. \cite{Chen2011,Mak2014}

In this article, we show that micro-Raman scattering measurements on electrochemically top-gated and SiO$_2$ back-gated graphene FETs can be used to accurately determine the geometrical capacitance of the electrical double layer and hence $E_{\rm F}$, with a spatial resolution down to approximately $ 1 ~\mu \rm m$. Calibrated electrochemical gates allow us (i) to quantitatively compare the anomalous doping-dependence of the G-mode phonon to theoretical models,\cite{Ando2006,Lazzeri2006,Pisana2007} (ii) to deduce the electron-phonon coupling constants at the center ($\Gamma$ point) and near the edges ($\rm K$ and $\rm K'$ points) of the Brillouin zone of graphene,\cite{Basko2009} and (iii) to establish well-defined correlations between the frequencies, linewidths and integrated intensities of the main Raman features in doped graphene. Importantly, we show that at top-gate voltages beyond the threshold for electrochemical reactions, defect concentrations of up to approximately $1.4\times10^{12}~\rm cm^{-2}$ can be created without damaging the device. This allows us, in particular, to quantitatively investigate the doping dependence of the defect-related D mode and to estimate the electron-defect scattering rate in graphene.

The paper is organized as follows: the experimental methods are exposed in Sec.~\ref{sec2}. Section~\ref{sec3} presents a model for the electric field effect in electrochemically-gated graphene FETs. Section~\ref{sec4} is dedicated to the experimental determination of the geometrical capacitance of the electrical double layer. In Sec.~\ref{secEPH} we specifically address electron-phonon coupling in pristine graphene. Section~\ref{sec8} describes our study of defective graphene and the determination of the electron-defect scattering rate. Finally, in Sec.~\ref{sec6}, we describe the correlations between the main Raman features in doped graphene. 

\begin{figure*}[!htb]
\begin{center}
\includegraphics[width=16.5cm]{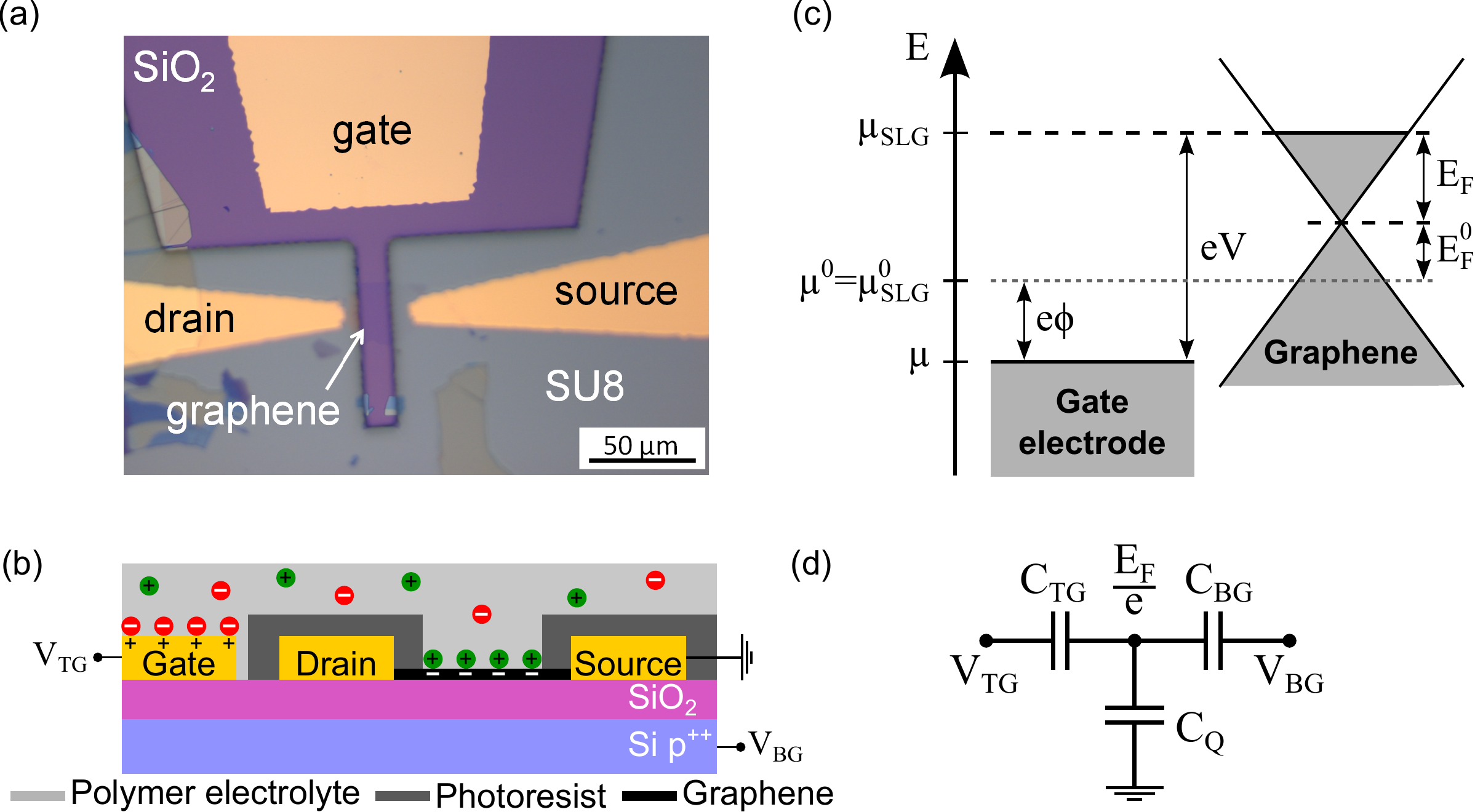}
\caption{(a) Optical image of a dual-gated graphene field-effect transistor prior deposition of the polymer electrolyte. The source and drain electrodes are covered with photoresist (SU8) to prevent them to be in contact with the polymer electrolyte. (b) A schematic cross-section of our dual gated graphene field-effect transistor, with Li$^+$ (green) and ClO$_4^-$ (red) ions and the electrical double layers near each electrode. The Si substrate is used as a back-gate. (c) Schematic energy diagrams of the electronic states of the gate electrode and of graphene. Occupied states are represented in grey. At zero gate voltage ($V=0$), the electrochemical potentials of the gate electrode $\mu$ and the graphene layer $\mu_{\textrm{SLG}}$ are equal. The Fermi energy of graphene is $E_{\textrm{F}}^0$. Applying a gate voltage $V$ results in  an electrostatically-induced shift $(e\phi)$ and in a change of the Fermi energy of graphene $(E_{\textrm{F}})$. The electrochemical potential difference is equal to $eV$, leading to Eq.~\eqref{eq_pot_chim} with $eV_0=E_{\textrm{F}}^0$. (d) Equivalent electrical circuit of our device at steady state. $C_{\rm BG}$ is the geometrical capacitance of the Si/SiO$_2$ back-gate, $C_{\rm G}$ is the geometrical capacitance of the electrical double layer at the graphene/polymer electrolyte interface and $C_{\rm Q}$ is the quantum capacitance of graphene.}
\label{Fig1}
\end{center}
\end{figure*}

\section{Methods}
\label{sec2}

\subsection{Sample preparation}
\label{sec2A}

Graphene samples are produced by mechanical exfoliation of natural graphite onto highly $p$-doped Si substrates covered with a ($285~\pm~15)~\textrm{nm}$ SiO$_2$ epilayer. Graphene monolayers are identified by optical microscopy and micro-Raman spectroscopy. Source, drain and gate electrodes are made by photolithography, followed by metal deposition (Ti (3 nm)/Au (47 nm)). The device are then coated with a $\sim$ 4 $\mu$m thick photoresist layer (MicroChem SU8 2005), and a second photolithography step is performed to open a window above the graphene channel and gate electrode,  as shown in Fig.~\ref{Fig1}(a)-(b). Finally, the electrochemical top-gate is formed by depositing a drop of polymer electrolyte with a micropipette. The polymer electrolyte is prepared by mixing lithium perchlorate (LiClO$_4$) and polyethylene oxide (PEO) in methanol at a weight ratio\cite{Das2008,Lu2004,Liu2013} 0.012:1:4. The mixture is then heated at 45 $^{\circ}$C and stirred until it becomes uniform. This suspension is filtered to get a clear solution. After dropcasting, the methanol evaporates and a thin film of transparent polymer electrolyte is formed. To remove residual moisture and solvent, the devices are annealed at about 90 $^{\circ}\textrm{C}$.
Noteworthy, the device geometry depicted in Fig.~\ref{Fig1}(a)-(b) features a well-defined gated region and prevents the polymer electrolyte to be in contact with the source and drain electrodes. As compared to earlier works,~\cite{Das2008,Das2009,Yan2009} this improves the gating efficiency and reduces the electrochemical reactivity of our devices. Some measurements described in the following are performed in a dual-gated geometry. In this case, the back-gate voltage is applied using the Si substrate as a gate electrode.

\subsection{Experimental setup}
\label{sec2B}

We perform micro-Raman scattering measurements in ambient conditions on top-gated and dual-gated graphene field-effect transistors. Raman spectra are recorded in a backscattering geometry, with a home-built setup, using a $\times$ 40 objective (NA = 0.60) and a 532~nm laser beam focused onto a spot of approximately $\sim1~\mu\textrm{m}$ in diameter. The sample holder is mounted onto a x-y-z piezoelectric stage, allowing spatially resolved Raman studies. The collected Raman scattered light is dispersed onto a charged-coupled device (CCD) array by a single-grating monochromator, with a spectral resolution of about $1~\textrm{cm}^{-1}$. The laser beam is linearly polarized and the laser power is maintained below 500~$\mu\textrm{W}$, in order to avoid thermally induced spectral shifts or lineshape modifications of the Raman features,\cite{Calizo2007} as well as photo-electrochemical reactions.\cite{Kalbac2010,Efetov2010,Bruna2014} The sample holder is electrically connected to a sourcemeter, which triggers our CCD array.  Raman spectra are recorded as a function of the applied gate bias, once a steady gate leak current (typically lower than 100~pA in the electrochemically top-gated configuration) is achieved. For this purpose, the gate bias is first applied for a settling time of $\sim$ 1~min, before recording each Raman spectrum.  This procedure ensures that Raman spectra are recorded at constant charge carrier densities. Raman spectra are also recorded during several forward and backward top-gate sweeps at the same spot on a given sample and very reproducible results, with no significant hysteresis, are observed. We find, however, that the geometrical capacitance of the top-gate, as well as the electron-phonon coupling constant may exhibit a certain degree of spatial inhomogeneity. Additionally, in ambient air, the gate capacitance may decrease over time, by up to one order of magnitude over a couple of days, due to a degradation of the polymer electrolyte. Such aging effects underscore the necessity of fast characterizations of electrochemically gated FETs and may account for the fairly large spread in the gate capacitances reported in literature.  In order to avoid sample aging effects, our measurements were performed immediately after deposition of the polymer electrolyte. Interestingly, the dispersions obtained from a set of measurements at several spots on a given graphene FET are very similar to the sample-to-sample dispersions observed by measuring at (single) random spots on a set of graphene FETs. This further highlights the interest of spatially resolved studies.

\section{Electric field effect}
\label{sec3}

\begin{figure*}[!tb]
\begin{center}
\includegraphics[width=17.8cm]{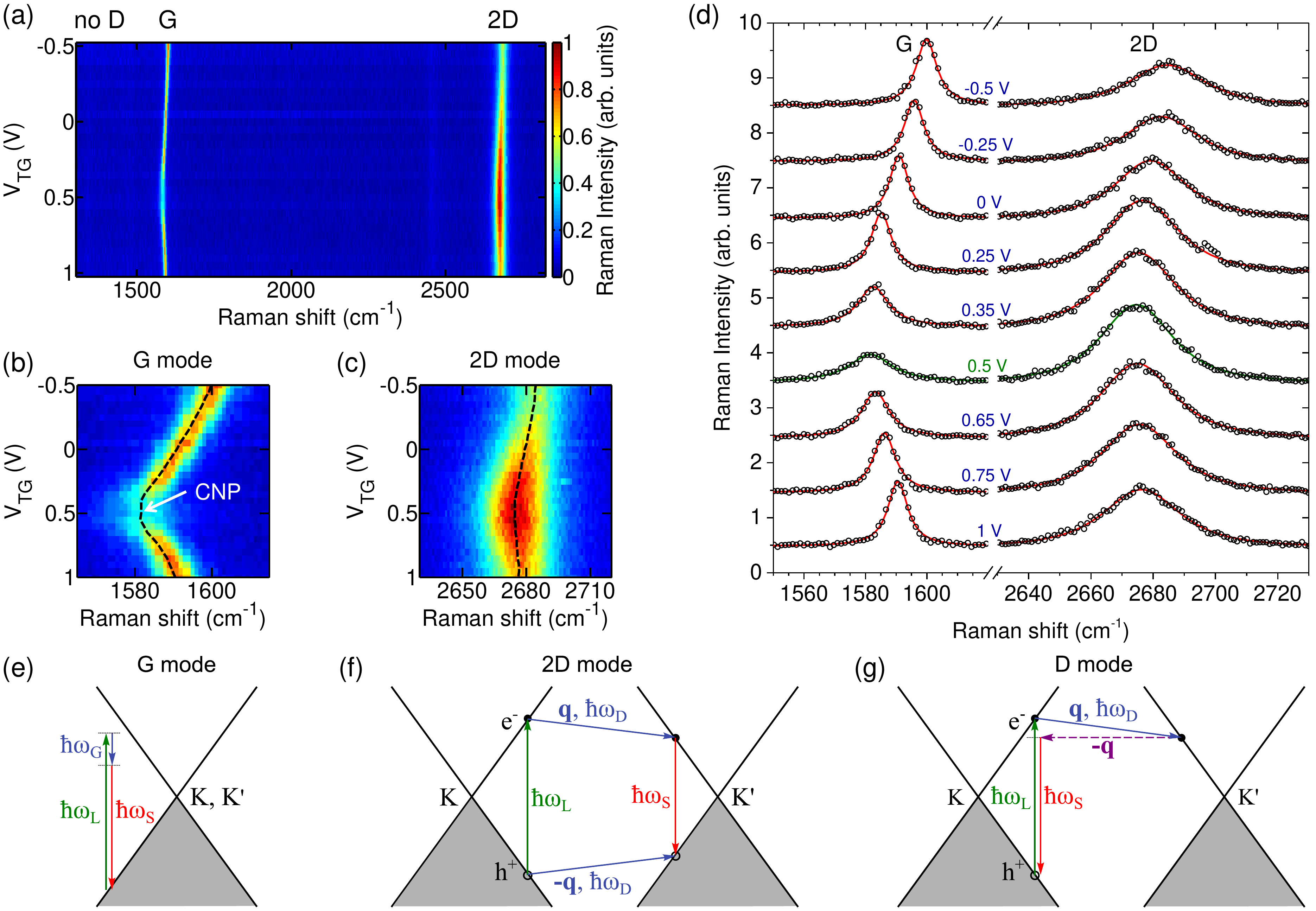}
\caption{(a)-(c) Color maps of the Raman spectra of a pristine graphene monolayer (sample 1), measured using a 532 nm laser beam, as a function of the top-gate voltage $V_{\rm TG}$. The G- and 2D-mode features appear prominently and no defect-induced D-mode feature is observed. Panels (b) and (c) show a clear evolution of the G- and 2D-mode features with varying $V_{\rm TG}$. The black dashed lines correspond to the central frequency of each Raman feature. The charge neutrality point (CNP) is indicated by an arrow. (d) Raman spectra at values of $V_{\rm TG}$ between -0.5~V and +1~V. The circles are the experimental data and the solid lines are fits (see text for details). The CNP is reached at $V_{\rm TG,0}=+0.5~\rm V$ (see green line). (e)-(g) Schematic representation, in the momentum-energy space, of the main Raman features in graphene.\cite{Ferrari2013} $\hbar \omega_{\rm L}$ ($\hbar \omega_{\rm S}$) denote the incoming laser (Raman scattered) photon energy. The G mode (e) is a non-resonant process.\cite{Basko2009,Chen2011} In (f), the 2D mode is represented as the dominant \textit{fully resonant inner process},\cite{Basko2008,Venezuela2011} involving two near zone-edge transverse optical phonons with frequency $\omega_{\rm D}$ and opposite momentum $\pm\mathbf{q}$. The defect-induced D mode \cite{Reich2000,Maultzsch2004,Venezuela2011,Ferrari2013} (g) involves one electron-phonon (solid arrow) scattering ($\mathbf{q}$, $\omega_{\rm D}$) and one electron-defect (dashed arrow) scattering process. One of these two processes is resonant. In (g), a resonant electron-phonon scattering process is represented.}
\label{Fig2}
\end{center}
\end{figure*}

Figure \ref{Fig2} shows typical Raman spectra recorded over a top-gate voltage sweep, with the two prominent Raman features in pristine graphene: the first order G-mode feature, which involves zone-center optical phonons (at the $\Gamma$ point), and the second-order resonant 2D-mode feature, which involves near zone-edge optical phonons (at the K and K' points).\cite{Malard2009,Ferrari2013} Note that no defect-induced D-mode feature emerges from the background in our experimental conditions. This illustrates the high structural quality of the graphene sample.
As expected,\cite{Das2008} the G-mode frequency and linewidth vary significantly with the top-gate bias ($V_{\rm TG}$). Similar trends are observed by applying a back-gate voltage ($V_{\rm BG}$). The minimum value of the G-mode frequency $\omega_{\rm G}$ and the maximum value of its full width at half maximum (FWHM) $\Gamma_{\rm G}$ are reached at the same value of $V_{\rm TG,0}=+0.5~\rm V$. This value corresponds to the charge neutrality point (CNP), where $E_{\rm F}=0$. The CNP is reached at a finite $V_{\rm TG,0}$,  due to an unintentional doping of the graphene layer, induced by the substrate as well as the polymer electrolyte.\cite{Das2008}
A finite value of $V_{\rm TG}-V_{\rm TG,0}$ results in a finite charge carrier density $n$. In this work, a positive (negative) gate voltage corresponds to electron (hole) injection.\footnote{Throughout the manuscript, $n$ will refer to the \textit{electron density}, such that positive (negative) $n$ correspond to electron (hole) doping.} Qualitatively, for both positive and negative values of $V_{\rm TG}-V_{\rm TG,0}$, we observe a nearly symmetric increase of  $\omega_{\rm G}$ accompanied by a symmetric decrease of $\Gamma_{\rm G}$ (see Sec.~\ref{sec5} for details). In contrast, the 2D-mode feature is less sensitive to doping than the G-mode feature~\cite{Das2008} (see Sec.~\ref{Sec2D}).

In order to carefully study the G- and 2D-mode features as a function of $E_{\rm F}$, one has to convert the gate voltage into $E_{\rm F}$ or, equivalently, $n$. First, the Fermi energy at a given $n$ is $E_{\rm F}=\sgn{ \left(n\right)} \hbar v_{\rm F}\sqrt{\pi |n|}$, where $\hbar$ is the reduced Planck's constant and $v_{F}\approx 1.1 \times 10^6~\rm m/s$ is the Fermi velocity of graphene on a SiO$_2$ substrate.\cite{Knox2008} Note that this formula applies only at $T=0$. However, in practice, finite temperature effects only induce a very minor correction to this simple scaling.\cite{Li2011}
An applied top- or back-gate voltage $V$ creates an electrostatic potential difference $\phi$ between the graphene monolayer and the gate electrode. Besides, the injection of charge carriers in graphene leads to a shift of its Fermi energy. Consequently, $V$ introduces a difference in the electrochemical potentials of the gate electrode $\mu$ and of the graphene layer $\mu_\textrm{SLG}$ (see Fig.~\ref{Fig1}(c))
\begin{equation}
\mu_{\textrm{SLG}}-\mu=eV=E_{\rm F}+e\phi+eV_{0},
\label{eq_pot_chim}
\end{equation}
where $e$ is the elementary charge, $V_{0}$ is a constant that accounts for the initial doping and implicitly includes the work function difference between the two materials.\cite{Giovannetti2008}

Assuming that the gate can be modeled as a parallel plate capacitor with a geometrical gate capacitance $C_{\rm G}$, the relation between $V$ and $E_{\rm F}$ is given by 
\begin{equation}
V-V_{0}=\frac{E_{\rm F}}{e}+\sgn(E_{\rm F})\frac{eE_{\rm F}^2}{\pi(\hbar v_{\rm F})^2C_{\rm G}}.
\label{eq_conv}
\end{equation}
Importantly, the first term on the right hand side of Eq.~\eqref{eq_conv} scales as $E_{\rm F}$ (\textit{i.e.,} $\sqrt{n}$) and is related to the quantum capacitance of graphene $C_\textrm{Q}$,\cite{Luryi1988} while the second term scales as $E_{\rm F}^2$ (\textit{i.e.,} as $n$) and is related to the geometrical gate capacitance $C_{\rm G}$. \cite{Das2008,Das2009}

For a typical SiO$_2$ back-gate insulator, the geometrical capacitance $C_{\rm BG}$ per unit area is simply given by $C_{\rm BG}=\varepsilon_r\varepsilon_0/d_{\rm BG}$, where $\varepsilon_r\approx4$ is the relative permittivity of SiO$_2$, $\varepsilon_0$ the vacuum permittivity and $d_{\textrm{BG}}$ is the SiO$_2$ thickness. In this work, $d_{\rm BG}=(285~\pm~15)~\textrm{nm}$ results in a back-gate capacitance $C_{\rm BG}=(12.4~\pm~0.7)~\textrm{nF~cm}^{-2}$. For a typical Fermi energy $E_{\rm F}\sim 100~\textrm{meV}$, the quantity $E_{\rm F}/e$ is negligible as compared to the other term in Eq.~\eqref{eq_conv}.

The case of the polymer electrolyte top-gate is slightly more complicated. Indeed, when a voltage is applied between the gate and the SLG, Li$^+$ and ClO$_4^-$ diffuse in the polymer to form electrical double layers at the interfaces as it is sketched in Fig.~\ref{Fig1}(b).\cite{Das2008} These EDL can be modeled as parallel plate capacitors with a thickness given by the Debye length $d_{\rm TG}$, and a geometrical capacitance per unit area $C_{\rm TG}=\varepsilon_r\varepsilon_0/d_{\rm TG}$. The total geometrical capacitance of the polymer electrolyte  is thus given by $C_{\rm TG}\left(S^{-1}_{\rm p-gate}+S^{-1}_{\rm p-graphene} \right)^{-1},$ where $S_{\rm p-gate}$ (resp. $S_{\rm p-graphene}$) is the contact area between the polymer electrolyte and the gate electrode (resp. the graphene monolayer). Since $S_{\rm p-gate} \gg S_{\rm p-graphene}$ (see Fig.~\ref{Fig1}(a)), one only needs to take into account the geometrical capacitance of the EDL at the graphene-polymer electrolyte interface. The Debye length is theoretically given by~\cite{Das2008} $d_{\rm TG}=2 \mathcal{C} e^2/\varepsilon_0\varepsilon_r k_{\rm B} T$, where $T$ is the temperature, $k_{\rm B}$ is Boltzmann's constant and $\mathcal{C}$ is the concentration of ions in the polymer electrolyte. In practice, the exact value of $\mathcal{C}$ cannot be measured. One can nevertheless obtain an estimate of $C_{\rm TG}\approx 4.4~\mu\textrm{F~cm}^{-2}$, assuming a typical value of $d_{\rm TG}\approx 1~\textrm{nm}$ and $\varepsilon_r \approx 5$ for PEO.\cite{Das2008} This capacitance is more than two orders of magnitude larger than $C_{\rm BG}$ and becomes comparable to the quantum capacitance for $E_\textrm{F}\sim 100~\textrm{meV}$. As a result, the two terms in Eq.~\eqref{eq_conv} are of the same order of magnitude and must be taken into account in the present study.

\section{Geometrical capacitance of the electrical double layer}
\label{sec4}

\begin{figure}[!tb]
\begin{center}
\includegraphics[width=8.35cm]{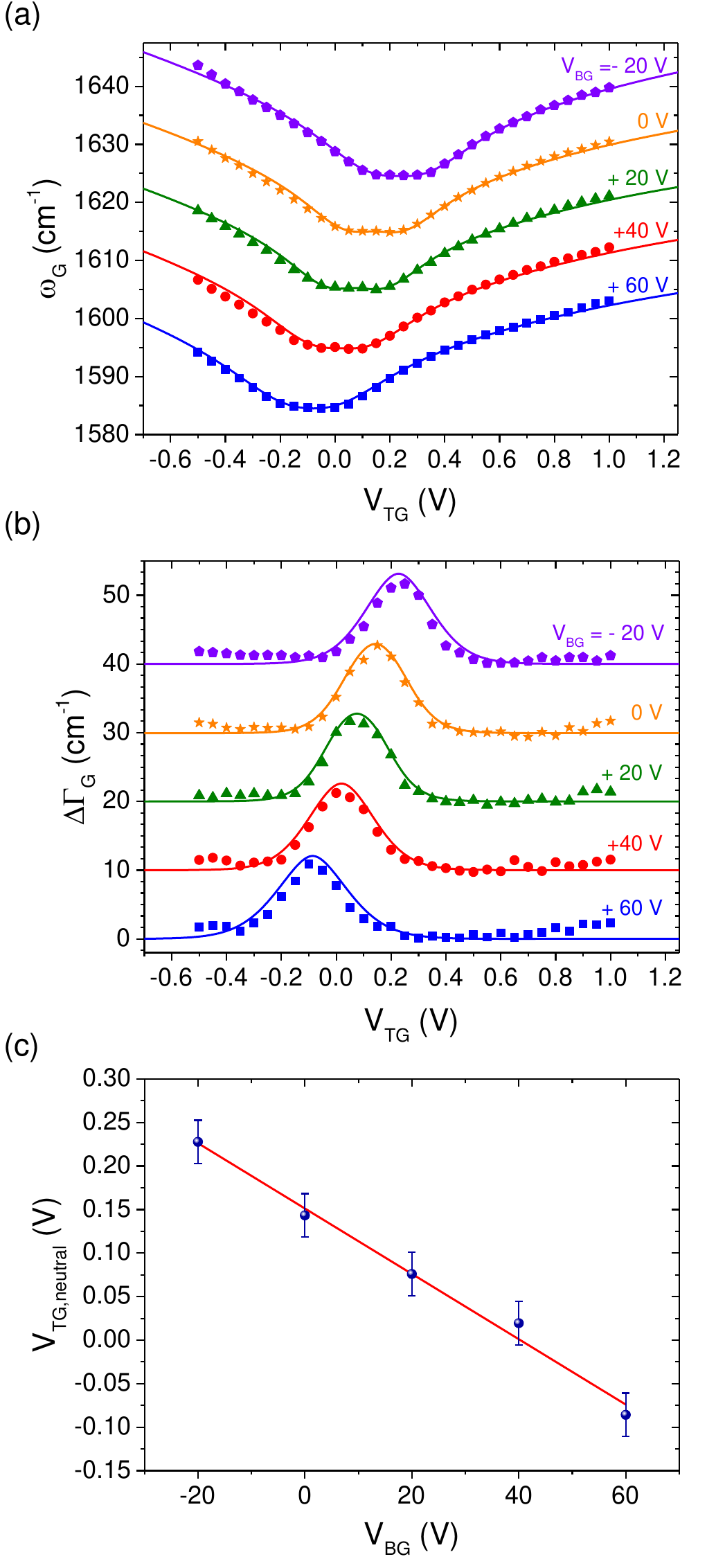}
\caption{(a) Frequency $\omega_{\rm G}$ and (b) relative FWHM $\Delta\Gamma_{\rm G}$ of the G-mode feature as a function of the top-gate voltage, recorded at various back-gate voltages on sample 1. The curves are vertically offset by 10 cm$^{-1}$ for clarity. The symbols are experimental data. (c) Top-gate voltage $V_{\rm TG,neutral}$, corresponding to the CNP in dual-gated graphene, as a function of the applied back-gate voltage $V_{\rm BG}$. A top gate capacitance $C_\textrm{TG} = 3.3~\mu\textrm{F~cm}^{-2}$ is deduced from a linear fit of the data (solid line). The solid lines in (a) and (b) are fits based on Eq.~\eqref{eq_fwhm} and \eqref{eq_posG}, respectively, with $C_\textrm{TG} = 3.3~\mu\textrm{F~cm}^{-2}$.}
\label{Fig3}
\end{center}
\end{figure}

Our first objective is to precisely determine $C_{\rm TG}$. Previous works on oxide dual-gated graphene FETs~\cite{Meric2008,Xu2011a} have shown that provided one geometrical capacitance is known, the other can be determined by monitoring the minimum (source-drain) conductivity point as a function of  the bottom and top-gate biases. At steady state, our dual-gated graphene FETs have the same equivalent electrical circuit (see Fig.~\ref{Fig1}(d)) as the devices of Ref.~\onlinecite{Xu2011a}. Here, rather than using electron transport measurements, we apply micro-Raman scattering spectroscopy, which provides a local measurement. For a fixed $V_{\rm BG}$, we sweep $V_{\rm TG}$ and record Raman spectra, as described in Sec.~\ref{sec2B}. Then, we extract $\omega_{\rm G}$ and $\Gamma_{\rm G}$ from Lorentzian fits. Figures \ref{Fig3}(a)-(b) show these two quantities as a function of $V_{\rm TG}$ for five different values of $V_{\rm BG}$. We observe a clear shift of the CNP, attained at $V_\textrm{TG,neutral}$, with $V_{\rm BG}$. In practice, $V_\textrm{{TG,neutral}}$ is extracted from the $\Gamma_{\rm G}(V_{\rm TG})$ curves, which, expectedly (see Sec.~\ref{sec5}), exhibit a sharper extremum near neutrality than the $\omega_{\rm G}(V_{\rm TG})$ curves. As shown in Fig.~\ref{Fig3}(c), $V_\textrm{TG,neutral}$ varies linearly with $V_{\rm BG}$. Indeed, from the equivalent circuit in Fig.~\ref{Fig1}(d), the total charge density injected by top- and back-gates leads~\cite{Xu2011a}
\begin{multline}
ne=-C_{\rm TG}\left(V_{\rm TG}-V_{\rm TG,0}-\frac{E_{\rm F}}{e}\right) \\
-C_{\rm BG}\left(V_{\rm BG}-V_{\rm BG,0}-\frac{E_{\rm F}}{e}\right).
\label{eq_charge}
\end{multline}
At the CNP, $n=0$ and $E_{\rm F}=0$. Therefore,
\begin{equation}
V_\textrm{TG,neutral}=V_{\rm TG,0}-\frac{C_{\rm BG}}{C_{\rm TG}}(V_{\rm BG}-V_{\rm BG,0}).
\label{eq_neutral}
\end{equation} 
Using Eq.~\eqref{eq_neutral}, a linear fit of the data in Fig.~\ref{Fig3}(c) yields $C_{\rm BG}/C_{\rm TG}=(3.8~\pm~0.2) \times 10^{-3}$. Since $C_{\rm BG}=(12.4~\pm~0.7)~\textrm{nF~cm}^{-2}$, we deduce that $C_{\rm TG}=(3.3~\pm~0.3)~\mu\textrm{F~cm}^{-2}$, which is of the same order of magnitude as what was reported before for similar devices.\cite{Das2008,Das2009,Shimotani2006,Efetov2010,Bruna2014} 
We may now convert $V_{\rm TG}$ into $E_{\rm F}$.

\section{Electron-phonon coupling in pristine graphene}
\label{secEPH}

\subsection{Doping-dependence of the G-mode feature}
\label{sec5}

\begin{figure}[!tb]
\begin{center}
\includegraphics[width=8.6cm]{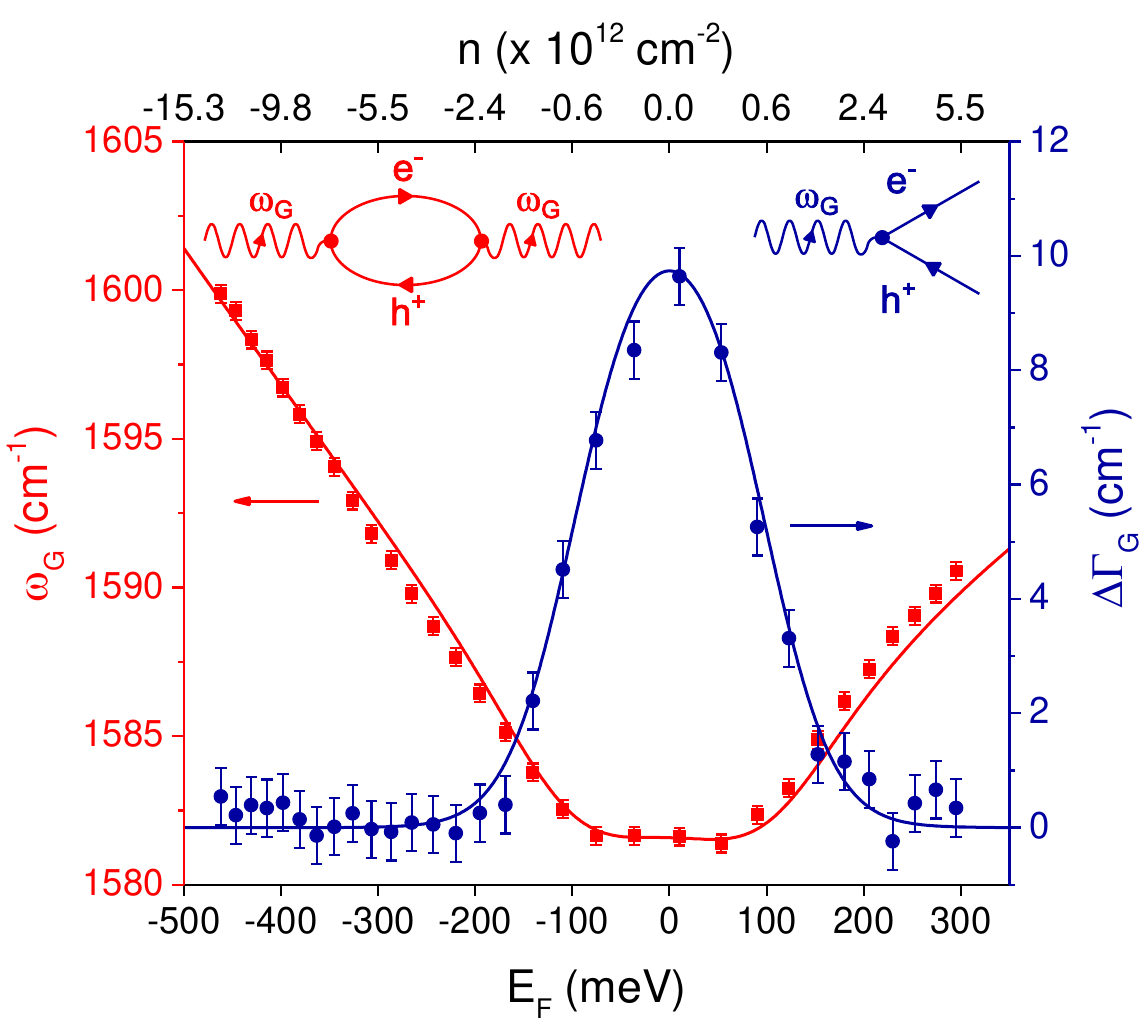}
\caption{Frequency $\omega_\textrm{G}$ (red squares, left axis) and relative FWHM $\Delta\Gamma_\textrm{G}$ (blue circles, right axis) of the G-mode feature, extracted from the measurements in Fig.~\ref{Fig2}, as a function of Fermi energy or doping. The corresponding Feynman diagrams are shown as insets. The left inset represents the renormalization of the G-mode phonon frequency due to interactions with virtual electron-hole pairs. The right inset represents lifetime broadening due to the resonant decay of a G-mode phonon into an electron-hole pair. The solid blue and red lines are fits based on Eqs.~\eqref{eq_fwhm} and \eqref{eq_posG}, respectively. The fitting parameters are $\textrm{C}_\textrm{TG} = 3.9~\mu\textrm{F~cm}^{-2}$, $\lambda_{\Gamma} = 0.027$ and $\delta E_\textrm{F} = 35~\textrm{meV}$.}
\label{Fig4}
\end{center}
\end{figure}

Considering only lattice expansion, due to the addition of charge carriers, one may expect the G-mode frequency to increase (decrease) under hole (electron) doping.\cite{Lazzeri2006} Thus, the peculiar, nearly symmetric behaviors observed here and previously reported by others~\cite{Pisana2007,Yan2007,Das2008,Kalbac2010,Chen2011,Chattrakun2013} contrast strongly with the trends predicted if one only considers lattice expansion effects. This \textit{anomalous} behavior has been originally predicted by Ando~\cite{Ando2006} and by Lazzeri and Mauri~\cite{Lazzeri2006} as a consequence of the strong coupling between zone-center optical phonons and low-energy electronic excitations across the gapless bands of graphene. Related effects occur in metallic carbon nanotubes.\cite{Piscanec2007} The anomalous doping dependence of the G-mode can be described using the phonon self-energy,\cite{Ando2006,Lazzeri2006,Pisana2007,Yan2007,Yan2008} the real part of which is equal to $\omega_{\rm G}$ and the imaginary part to $\Gamma_{\rm G}$. As a result, the evolution of $\omega_{\rm G}$ and $\Gamma_{\rm G}$ are deeply connected (see Fig.~\ref{Fig4}).

The variation of $\Gamma_{\rm G}$ is due to the decay of the G phonon into an electron-hole pair (see right inset in Fig.~\ref{Fig4}) and is given by
\begin{multline}
\Delta\Gamma_{\rm G}=\Gamma_{\rm G}-\Gamma_0\\
=\frac{\lambda_\Gamma}{4}\omega_{\rm G}^0\times \left[f\left(-\frac{\hbar\omega_{\rm G}^0}{2}-E_{\rm F}\right)-f\left(\frac{\hbar\omega_{\rm G}^0}{2}-E_{\rm F}\right)\right],
\label{eq_fwhm}
\end{multline} 
where $\omega_{\rm G}^0$ is the phonon frequency at $E_{\rm F}=0$, $f(E)=[1+\exp(E/k_BT)]^{-1}$ is the Fermi-Dirac distribution at a temperature $T$ and $\lambda_\Gamma$ is a dimensionless coefficient corresponding to the electron-phonon coupling strength\footnote{Here we choose to use the coupling constant $\lambda_\Gamma$ as defined by Basko in Ref.~\onlinecite{Basko2008}. In Refs.~\onlinecite{Lazzeri2006,Pisana2007} the dimensionless electron-phonon coupling constant is denoted $\alpha'$ and is defined as $\alpha'=\lambda_\Gamma / 2\pi$} (see also Sec.~\ref{sec7}). $\Gamma_0$ contains all other sources of broadening that are independent on the carrier density (anharmonic coupling,\cite{Bonini2007} disorder, instrument response function). For $\abs{E_{\rm F}}>\hbar\omega_{\rm G}^0/2$, $\Delta\Gamma_{\rm G}$ vanishes due to Pauli blocking.

The evolution of $\omega_{\rm G}$ with $E_{\rm F}$ is the sum of an \textit{adiabatic} contribution $\omega_{\rm G}^{\rm A}$ corresponding to the modification of the equilibrium lattice parameter and a \textit{non adiabatic} one $\omega_{\rm G}^{\rm NA}$ corresponding to the renormalization of the G-mode phonon energy due to interactions with virtual electron-hole pairs~\cite{Ando2006,Lazzeri2006} (see left inset in Fig.~\ref{Fig4}). At a finite temperature $T$, the frequency shift is given by\cite{Lazzeri2006,Pisana2007}
\begin{equation}
\Delta\omega_{\rm G} = \omega_{\rm G}-\omega_{\rm G}^0 = \Delta\omega_{\rm G}^{\rm A} + \Delta\omega_{\rm G}^{\rm NA} , \label{eq_posG}
\end{equation}
with
\begin{equation}
\Delta\omega_{\rm G}^{\rm NA} = \frac{\lambda_\Gamma}{2\pi \hbar}\dashint_{-\infty}^{+\infty} \frac{[f(E-E_{\rm F})-f(E)]E^2\sgn(E)}{E^2-(\hbar\omega_{\rm G}^0)^2/4} \, \mathrm{d}E, \label{eq_non_adia}
\end{equation}
where $\dashint$ denotes the Cauchy principal value. One should note that $\Delta\Gamma_{\rm G}$ and $\Delta\omega_{\rm G}$ are proportional to $\lambda_\Gamma$.

In the simulations described below, we use the results of the calculation by Lazzeri and Mauri to include \textit{adiabatic} contribution $\Delta\omega_\textrm{G}^\textrm{A}$ (see Eq.~(3) in Ref.~\onlinecite{Lazzeri2006}). Importantly, for $\abs{E_{\rm F}}<1~\rm{eV}$, the adiabatic contribution provides only a minor correction to the non-adiabatic term and does not affect $\Gamma_{\rm G}$. 

Moreover, to accurately describe the experimental evolution of the G mode, one also has to take into account random spatial fluctuations of the Fermi energy.\cite{Casiraghi2007,Martin2008,Xu2011b,Li2011} It is reasonable to assume that $E_{\rm F}$ follows a Gaussian distribution~\cite{Martin2008,Xu2011b,Li2011}  around its mean value, with a standard deviation $\delta E_{\rm F}$. Thereafter, the computed $\Delta\omega_{\rm G} (E_{\rm F})$ and $\Delta\Gamma_{\rm G} (E_{\rm F})$ used to fit our data are given by the convolution of this Gaussian distribution with Eq.~\eqref{eq_fwhm}~and~\eqref{eq_posG}.

Figure~\ref{Fig3} displays the results of \textit{simultaneous} fits of $\Delta\omega_{\rm G} (V_{\rm TG})$ and $\Delta\Gamma_{\rm G} (V_{\rm TG})$ for five top-gate sweeps at different $V_{\rm BG}$. We used $v_{\rm F}=1.1 \times 10^6~\textrm{m~s}^{-1}$ and the values of $C_{\rm TG}$, $V_\textrm{TG,neutral}$ and $\omega_{\rm G}^0$ obtained in Sec.~\ref{sec4}. Thus the fitting parameters are $\lambda_\Gamma$, $\delta E_{\rm F}$ and $\Gamma_0$.
The experimental data are remarkably well fitted by the theoretical model. Interestingly, although the two phonon anomalies\cite{Ando2006,Lazzeri2006,Yan2008} predicted at $E_{\rm F}=\pm \hbar\omega_G^0$ by Eq.~\eqref{eq_posG} are largely smeared out at room temperature, one can still notice a hint of their presence in Fig.~\ref{Fig3}(a), \ref{Fig4} and \ref{Fig9}(a).

From these five fits, we get $\lambda_\Gamma=0.036 $ and $\delta E_{\rm F}= 40~\textrm{meV}$. Since $\delta E_{\rm F}\approx50~\textrm{meV}$ on bare SiO$_2$ without an electrochemical top-gate,\cite{Martin2008,Xue2011} we conclude that charge inhomogeneity does not have a major effect on our analysis. DFT calculations~\cite{Lazzeri2006,Pisana2007} have predicted $\lambda_\Gamma=0.028$, which is slightly smaller, but consistent with our measurement. 

Another way to further compare the experimental data and theory is to set $C_{\rm TG}$ as adjustable parameter when fitting $\Delta\omega_{\rm G} (V_{\rm TG})$ and $\Delta\Gamma_{\rm G} (V_{\rm TG})$. This yields $C_{\rm TG}=3.9~\mu\textrm{F~cm}^{-2}$, $\lambda_\Gamma=0.034$ and $\delta E_{\rm F}= 35~\textrm{meV}$. These values are very consistent with the more constrained fits discussed above (see Sec.~\ref{sec4}). Similar studies were repeated on more than five samples, with similar conclusions. This demonstrates that a direct fit of $\Delta\omega_{\rm G}(V_{\rm TG})$ and $\Delta\Gamma_{\rm G}(V_{\rm TG})$ can be used to get an accurate measurement of $C_{\rm TG}$, which allows to convert $V_{\rm TG}$ into $E_{\rm F}$ through Eq.~\eqref{eq_conv}.  This is a much faster approach to determine $C_{\rm TG}$, which does not require a dual-gated device. As an example, a fit of the data in Fig.~\ref{Fig2} is shown in Fig.~\ref{Fig4}, and shows a very good agreement between experiment and theory. More generally, our fitting procedure allows us to estimate $C_{\rm TG}$ and $\lambda_{\Gamma}$ with relative uncertainties of approximately $20\%$ and $10\%$, respectively.

To better understand the importance to fit simultaneously $\Delta\omega_{\rm G}(V_{\rm TG})$ and $\Delta\Gamma_{\rm G}(V_{\rm TG})$, we have fit these quantities separately for the measurements shown in Fig.~\ref{Fig3} (not shown). From the fit of $\Delta\omega_{\rm G}(V_{\rm TG})$, we obtain $C_{\rm TG}=2.3~\mu\textrm{F~cm}^{-2}$, $\lambda_\Gamma=0.042$ and $\delta E_{\rm F}=50~\textrm{meV}$. Except the large value of $\lambda_\Gamma$, the two parameters are reasonable. From the fit of $\Delta\Gamma_{\rm G}$, we obtain $\lambda_\Gamma=0.033$ and $\delta E_{\rm F}\approx40~\textrm{meV}$ and an unrealistically large $C_{\rm TG}\sim100~\mu\textrm{F~cm}^{-2}$. The latter value suggests that the behavior of $\Delta\Gamma_{\rm G}$ can be rationalized using solely the quantum capacitance of graphene. This is understandable, since the variations of $\Delta\Gamma_{\rm G}$ occur near $E_{\rm F}=0$, where the contribution of the quantum capacitance dominates in Eq.~\eqref{eq_conv}. However, the value of $\Delta\Gamma_{\rm G}$ near $E_{\rm F}=0$ is directly proportional to $\lambda_\Gamma$ and is not influenced by $C_{\rm TG}^{}$, while $\Delta\omega_{\rm G}$ varies mostly away from the CNP. Hence, its evolution with $V_{\rm TG}$ is influenced by both $\lambda_\Gamma$ and $C_{\rm TG}$.
Consequently, a simultaneous fit allows for a reliable estimation of $\lambda_\Gamma$ (through the doping dependence of $\Delta\Gamma_{\rm G}(V_{\rm TG})$), and, in turn of $C_{\rm TG}$ (through the slope of $\Delta\omega_{\rm G}(V_{\rm TG})$ curve, having $\lambda_\Gamma$ constrained by $\Delta\Gamma_{\rm G}(V_{\rm TG})$).

\begin{figure}[!tb]
\begin{center}
\includegraphics[width=8.6cm]{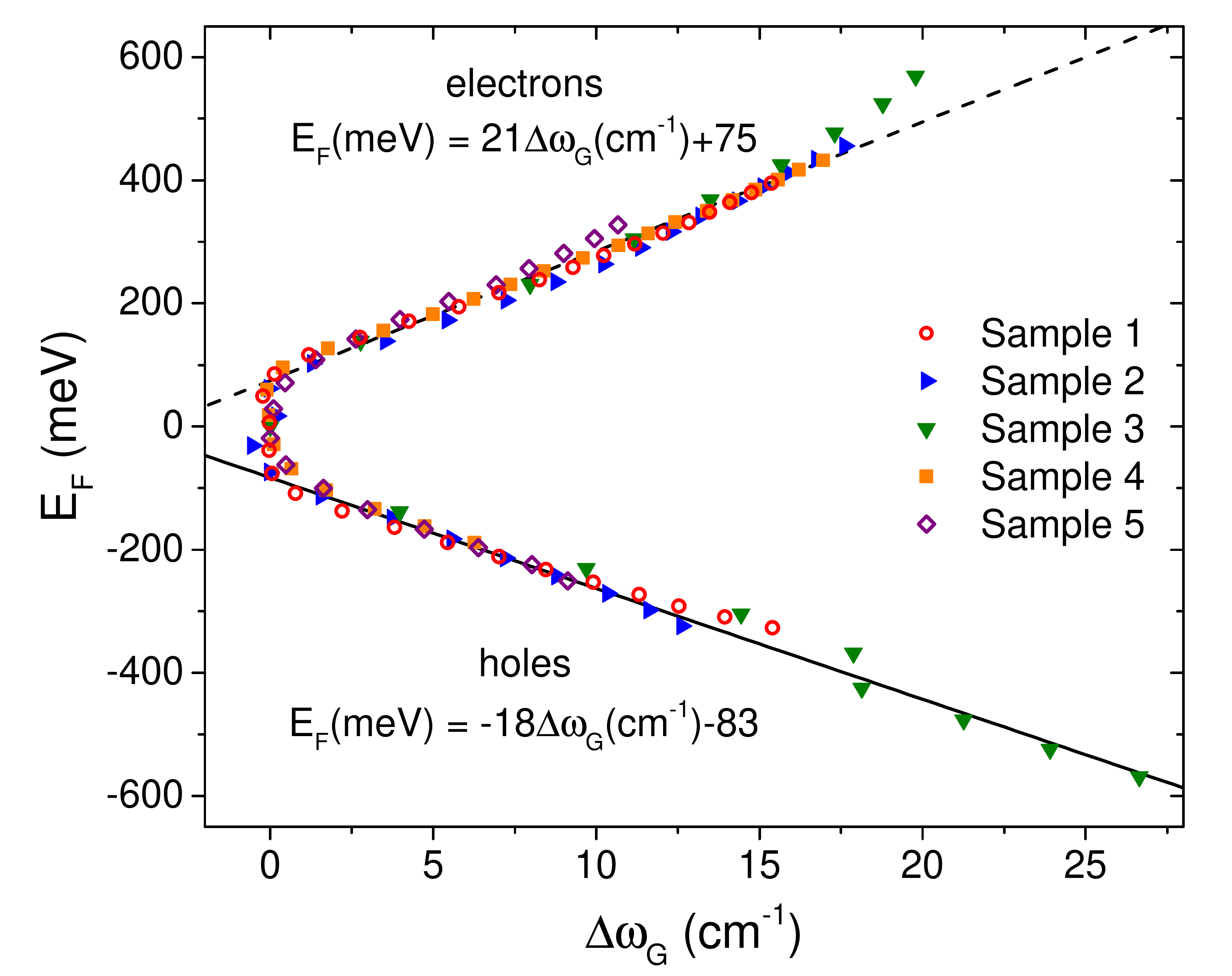}
\caption{Fermi energy $E_{\rm F}$ as a function of the relative frequency of the G mode $\Delta\omega_{\rm G}$. Measurements on five different devices are represented with different symbols. The dashed and solid lines correspond to Eq.~\eqref{eq_elec} and \eqref{eq_trou}, respectively.}
\label{Fig5}
\end{center}
\end{figure}

Figure \ref{Fig5} shows the evolution of $E_\textrm{F}$ as a function of $\Delta\omega_\textrm{G}$ for five different graphene FETs (denoted sample 1 to 5) in which $C_\textrm{TG}$ and $\lambda_\Gamma$ have been previously determined by the \textit{simultaneous} fit of $\Delta\omega_\textrm{G}(V_{\rm TG})$ and $\Delta\Gamma_\textrm{G}(V_{\rm TG})$. For these five samples, we found an average of $\left<C_\textrm{TG}\right>=(4.5~\pm~1.5)~\mu\textrm{F~cm}^{-2}$ and $\left<\lambda_\Gamma\right>=(0.032~\pm~0.004)$ (see also Table~\ref{table1}). 
This translates into an average relative G-mode FWHM (see Eq.~\eqref{eq_fwhm}) of $\Delta\Gamma_\textrm{G}=12.6~\pm~1.6~\textrm{cm}^{-1}$ (at $T=0$, $E_{\rm F}=0$ and $\delta E_{\rm F}=0$) that is consistent with the value of $\Gamma_\textrm{G}\approx 15~\textrm{cm}^{-1}$ recorded on quasi-undoped suspended graphene at low temperature.\cite{Berciaud2013}
Remarkably, and in spite of the different values of $C_{\rm TG}$, the data for these five devices shown in Fig.~\ref{Fig5} collapse onto a same curve. In practice, this very reproducible behavior can be used to evaluate $E_{\rm F}$ knowing $\Delta\omega_{\rm G}$, which is of broad interest in graphene science. For this purpose, we consider the asymptotic behavior of $\Delta\omega_{\rm G}(E_{\rm F})$. When $\abs{E_{\rm F}}\gg\hbar\omega_{\rm G}^0/2$, Eq.~\eqref{eq_non_adia} becomes 

\begin{equation}
\Delta\omega_{\rm G}^{\rm NA}\approx\frac{\lambda_\Gamma}{2\pi \hbar}\abs{E_{\rm F}}.
\label{eq_asymp}
\end{equation}
Assuming that the adiabatic contribution $\Delta\omega_{\rm G}^{\rm A}$ is negligible compared to $\Delta\omega_{\rm G}^{\rm NA}$, $\Delta\omega_{\rm G}$ should be linear with $\abs{E_{\rm F}}$. Indeed, in Fig.~\ref{Fig5} for the five different samples, $\omega_{\rm G}(E_{\rm F})$ clearly scales linearly for $\abs{E_{\rm F}}\gtrsim 100~\textrm{meV}$. The slightly different slopes observed for electron and hole doping arise from the opposite sign of the adiabatic corrections.

For $\abs{E_{\rm F}}\gtrsim 100~\textrm{meV}$, we find
\begin{align}
E_{\rm F} &\gtrsim +100~\textrm{meV},&~ E_{\rm F}=+21\Delta\omega_{\rm G} + 75, \label{eq_elec}\\
E_{\rm F} &\lesssim -100~\textrm{meV},&~ E_{\rm F}=-18\Delta\omega_{\rm G} -83, \label{eq_trou}
\end{align}
where $E_{\rm F}$ is expressed in meV and $\Delta\omega_{\rm G}$ in $\textrm{cm}^{-1}$. However, it should be noted that this linear scaling only holds for $\abs{E_{\rm F}}\lesssim 500-600~\textrm{meV}$. In fact, for higher $\abs{E_{\rm F}}$, $\Delta\omega_{\rm G}$ no longer scales linearly with $\abs{E_{\rm F}}$ since $\Delta\omega_{\rm G}^{\rm A}$ can no longer be neglected compared to $\Delta\omega_{\rm G}^{\rm NA}$. Moreover, Eqs.~\eqref{eq_elec} and \eqref{eq_trou} can be applied provided the shift in $\Delta\omega_{\rm G}$ is exclusively due to doping, \textit{i.e.,} other extrinsic factors, such as mechanical strain do not contribute. If this is not the case, one has to separate the various contributions, using, \textit{e.g.}\ the method described in Ref.~\onlinecite{Lee2012} with the results of Sec.~\ref{sec6B}.

\subsection{Doping-dependence of the 2D-mode feature}
\label{Sec2D}
\begin{figure}[!tb]
\begin{center}
\includegraphics[width=8.6cm]{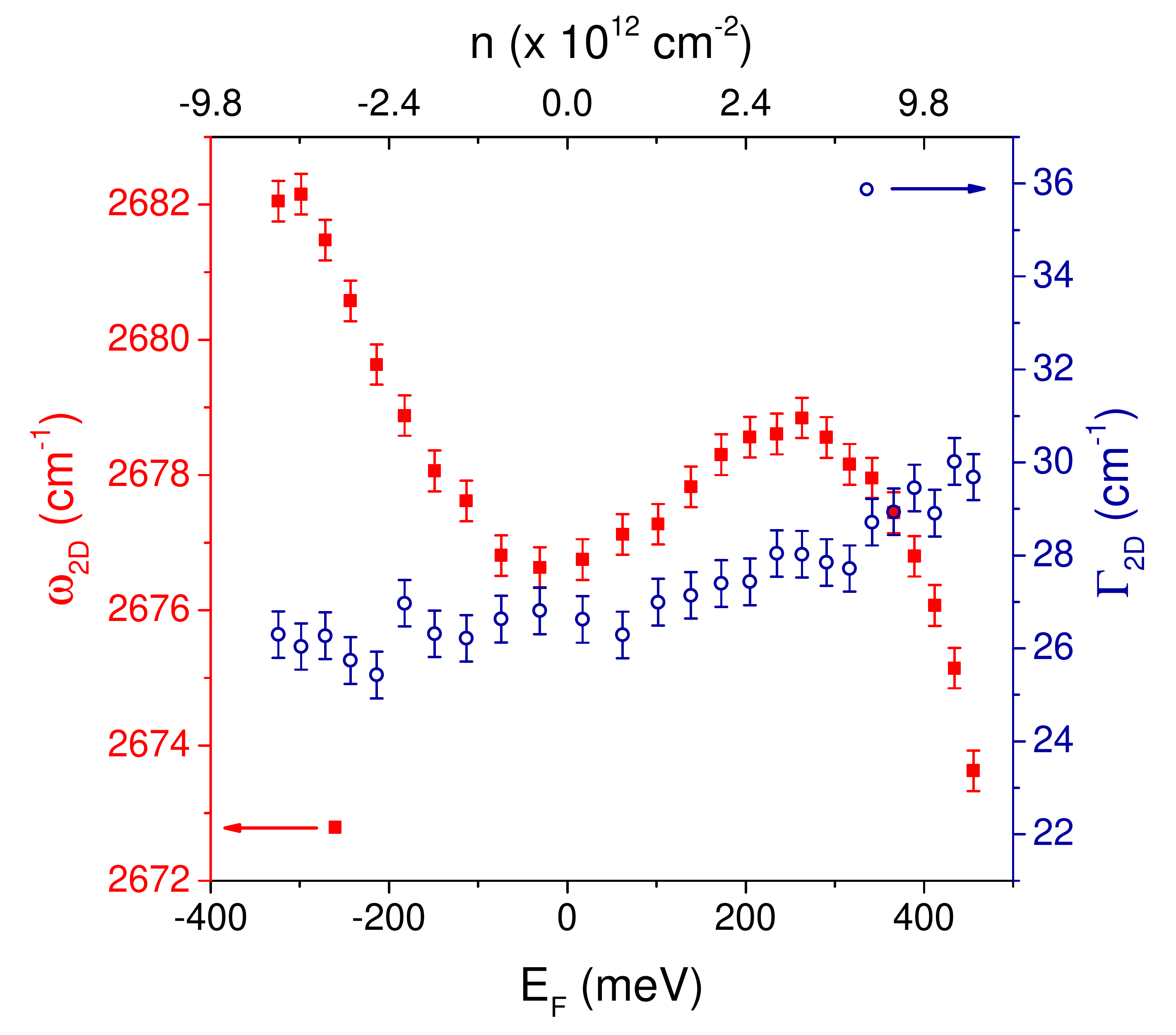}
\caption{Doping dependence of the frequency $\omega_{\rm 2D}^{}$ (red squares, left axis) and FWHM $\Gamma_{\rm 2D}^{}$ (blue circle, right axis) of the 2D mode-feature. The measurements are performed on sample 2, before the creation of defects.}
\label{Fig6}
\end{center}
\end{figure}

Let us briefly comment on the 2D-mode feature. Figure \ref{Fig6} shows the evolution of the frequency $\omega_{\rm 2D}(E_{\rm F})$ and FWHM $\Gamma_{\rm 2D}(E_{\rm F})$ of the 2D-mode feature with $E_{\rm F}$ for sample 2 (2D mode spectra are also shown for sample 1 in Fig.~\ref{Fig2}). In supported graphene, the 2D-mode feature typically exhibit a quasi-symmetric lineshape that can be phenomenologically fit to a modified Lorentzian profile.\cite{Basko2008,Berciaud2013} We find that $\Gamma_{\rm 2D}(E_{\rm F})$  does not vary significantly with the gate bias, while $\omega_{\rm 2D}$ varies little at moderate doping $(\abs{E_{\rm F}}\lesssim200~\rm meV)$, but tends to stiffen (soften) significantly for stronger hole (electron) doping. The observed evolution of $\omega_{\rm 2D}$ outlined in Fig.~\ref{Fig6} (see also Figs.~\ref{Fig2} and \ref{Fig9}) can be qualitatively understood as the sum of a dominant adiabatic contribution and a weaker non-adiabatic contribution. The latter is reduced as compared to the case of the G-mode feature, likely because the 2D-mode feature involves phonons that are significantly away from the edges of the Brillouin zone.\cite{Das2008}

\subsection{Electron-electron and electron-phonon scattering}
\label{sec7}

\begin{figure}[!tb]
\begin{center}
\includegraphics[width=8.6cm]{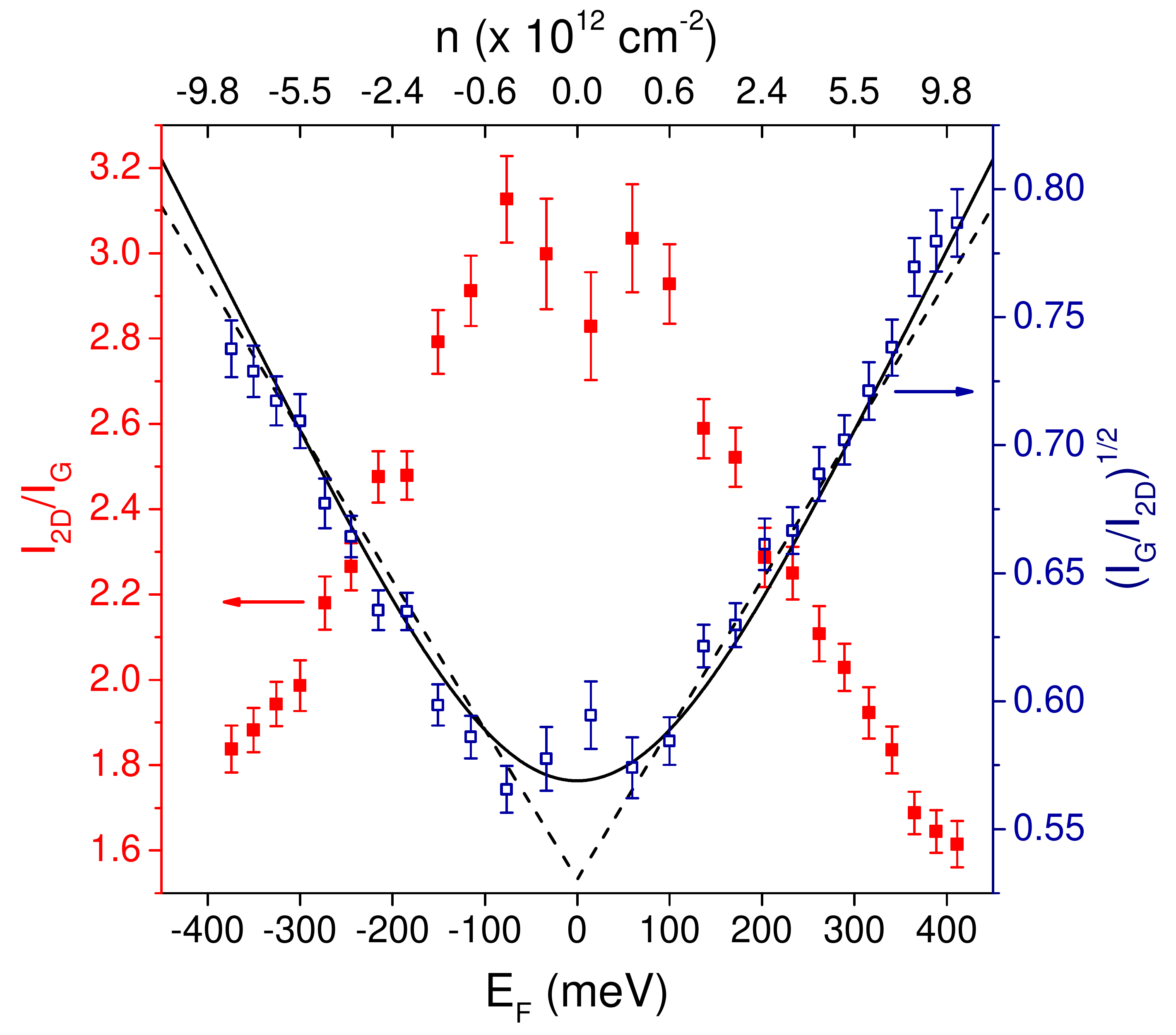}
\caption{Left axis: doping dependence of the ratio between the integrated intensity of the 2D-mode feature and that of the G-mode feature. Right axis: doping dependence of the square root of the ratio between the integrated intensity of the G mode and that of the 2D mode.  The dashed and solid lines are fits based on Eq.~\eqref{eq_ratio}, without and with Fermi energy fluctuations, respectively. A value of $\gamma_{\textrm{e-ph}}=51~\textrm{meV}$ is deduced from the fit. The measurements are performed on sample 2, before the creation of defects.}
\label{Fig7}
\end{center}
\end{figure}

Another useful quantity is the integrated intensity of a Raman feature (denoted $I_{\rm X}$), which represents the total probability of the Raman scattering process. The integrated intensity of the 2D-mode feature ($I_{\rm 2D}$) depends on $E_{\rm F}$,\cite{Das2008,Das2009,Basko2009} whereas $I_{\rm G}$ does not, as long as $\abs{E_{\rm F}}\leq \hbar\omega_{\rm L}/2$, where $\hbar\omega_{\rm L}$ is the energy of the incident laser.\cite{Basko2009b,Kalbac2010,Chen2011} In Fig.~\ref{Fig7}, we consider the ratio $I_{\rm 2D}/I_{\rm G}$, which is maximum for $E_{\rm F}=0$ and decreases almost symmetrically for increasing $\abs{E_{\rm F}}$. Following Ref.~\onlinecite{Basko2008,Basko2009}, the integrated intensity of the 2D-mode feature writes

\begin{equation}
I_{\rm 2D}\propto \left(\frac{\gamma_{\rm K}}{\gamma_{\rm e-ph}+\gamma_{\rm D}+\gamma_{\rm ee}}\right)^2,
\label{eqI2D}
\end{equation}
where $\gamma_{\rm e-ph}+\gamma_{\rm D}+\gamma_{\rm ee}$ is the total electron scattering rate, with $\gamma_{\rm e-ph}$ the electron-phonon scattering rate, $\gamma_{\rm D}$ the electron-defect scattering rate, and $\gamma_{\rm ee}$ the electron-electron scattering rate. The electron-phonon scattering rate can be approximated as $\gamma_{\rm e-ph}=\gamma_{\rm K}+\gamma_{\Gamma}$, where  $\gamma_{\rm K}$ and $\gamma_{\Gamma}$ are the scattering rates for zone-edge and zone-center optical phonons, respectively. Note that Eq.~\eqref{eqI2D} is obtained under the assumption of a \textit{fully resonant} process (see Fig.~\ref{Fig2}(f)), and that trigonal warping effects leading to momentum-dependent scattering rates are neglected.\cite{Basko2008,Basko2009,Basko2013,Venezuela2011}
While $\gamma_{\rm D}$ and $\gamma_\textrm{e-ph}$ do not depend on $E_{\rm F}$, $\gamma_{\rm ee}$ has been predicted to scale linearly with $\abs{E_{\rm F}}$. For $\abs{E_{\rm F}}\ll \hbar\omega_{\rm L}/2$, Basko \textit{et al.} calculated\cite{Basko2009}
\begin{equation}
\sqrt{\frac{I_{\rm G}}{I_{\rm 2D}}}=\frac{\left.\sqrt{\frac{I_{\rm G}}{I_{\rm 2D}}}\right|_0}{\gamma_{\textrm{e-ph}}+\gamma_{\rm D}}(\gamma_{\textrm{e-ph}}+\gamma_{\rm D}+0.06\abs{E_{\rm F}}),
\label{eq_ratio}
\end{equation}

where $\left.\sqrt{\frac{I_{\rm G}}{I_{\rm 2D}}}\right|_0$ corresponds to the value at $E_{\rm F}=0$.

In this section, we are considering pristine graphene, in which $\gamma_{\rm D}\ll\gamma_{\rm e-ph}$. As illustrated by the dashed line in Fig.~\ref{Fig7}, our experimental data agree well with a fit based on Eq.~\eqref{eq_ratio} for $\abs{E_{\rm F}}\gtrsim 100~\textrm{meV}$. However, we observe a deviation from Eq.~\eqref{eq_ratio} near the CNP, likely due to Fermi energy fluctuations. As in Sec.~\ref{sec5}, we therefore fit the experimental data with the Gaussian convolution of Eq.~\eqref{eq_ratio}, resulting in the solid line in Fig.~\ref{Fig7}. The agreement between theory and experiment is very good and more compelling than in the seminal study in Ref.~\onlinecite{Basko2009}. The fitting parameters are  $\left.\frac{I_{\rm 2D}}{I_{\rm G}}\right|_0 = 3.6$, $\gamma_{\textrm{e-ph}}=51~\textrm{meV}$ and  $\delta E_{\rm F}= 110~\textrm{meV}$. 
We repeated this analysis on three pristine samples and found average values of $\left<\gamma_{\textrm{e-ph}}\right>=(47~\pm~7)~\textrm{meV}$, $\left<\delta E_{\rm F}\right>= (120~\pm~10)~\textrm{meV}$ and  $\left<\left.\frac{I_{\rm 2D}}{I_{\rm G}}\right|_0\right> = 4.2~\pm~0.6$ (see Table~\ref{table1}). Note that the dispersion of the measurements on these three devices is very similar to the dispersion observed when measuring on several spots on the same sample. The value of $\gamma_{\textrm{e-ph}}$ is in good agreement with the estimate in Ref.~\onlinecite{Basko2009}. The Fermi energy fluctuation $\delta E_{\rm F}$ obtained here is more realistic than the lower values estimated from the simultaneous fit of $\Delta\omega_{\rm G}(V_{\rm TG})$ and $\Delta\Gamma_{\rm G}(V_{\rm TG})$ (see Sec.~\ref{sec5}). It corresponds to a charge inhomogeneity of $\delta n \lesssim 10^{12}~\textrm{cm}^{-2}$, in line with previous scanning tunneling microscopy measurements.\cite{Martin2008,Xue2011}

Interestingly, in Ref.~\onlinecite{Basko2009}, the authors claim that the \textit{intrinsic} value of $\left.\frac{I_{\rm 2D}}{I_{\rm G}}\right|_0$ for undoped graphene is in the range 12-17 (using a 514.5~nm excitation wavelength). However, this estimation is based on Raman measurements on quasi-undoped suspended graphene\cite{Berciaud2009} and does not take into account the effect of optical interferences, which occur in graphene-based multilayer structures and may critically affect the intensity of the Raman features.\cite{Yoon2009,Metten2014}
From the data in Ref.~\onlinecite{Metten2014}, an \textit{intrinsic} value corrected from interference effects of $\left.\frac{I_{\rm 2D}}{I_{\rm G}}\right|_\textrm{intr} = 5~\pm~0.3 $ can be estimated for freely suspended, undoped graphene, using a 532~nm excitation wavelength, as in the present study. Considering the distinct Raman enhancement factors for the G- and 2D-mode features in the PEO/graphene/SiO$_2~(285~\rm nm)$/Si multilayer structure, our average value of $\left<\left.\frac{I_{\rm 2D}}{I_{\rm G}}\right|_0\right> = 4.2~\pm~0.6$ translates into an average \textit{intrinsic} value of $\left<\left.\frac{I_{\rm 2D}}{I_{\rm G}}\right|_\textrm{intr}\right> = 4.9~\pm~0.7 $ (see Table~\ref{table1}), which is in excellent agreement with our estimate on suspended graphene.

As outlined in Ref.~\onlinecite{Basko2008,Basko2009,Basko2008b}, the scattering rate $\gamma_{\textrm{e-ph}}$ is linked to the dimensionless electron-phonon coupling constants $\lambda_{\Gamma}$ and $\lambda_K$ through
\begin{equation}
\gamma_{\textrm{e-ph}}=\gamma_{\rm K}+\gamma_{\Gamma}=\frac{\lambda_{\rm K}}{4}\left(\frac{\hbar\omega_{\rm L}}{2}-\hbar\omega_{\rm K}\right)+\frac{\lambda_{\Gamma}}{4}\left(\frac{\hbar\omega_{\rm L}}{2}-\hbar\omega_{\Gamma}\right),
\label{eq_scattering}
\end{equation}
where $\hbar\omega_{\rm K}\approx 1210~\textrm{cm}^{-1}= 150~\textrm{meV}$ is the in-plane transverse optical (TO) phonon energy at the K (K') point, $\hbar\omega_{\Gamma}:=\hbar\omega_{\rm G}\approx 1580~\textrm{cm}^{-1}=196~\textrm{meV}$ is the in-plane optical phonon energy at $\Gamma$ (\textit{i.e.}, the G-mode frequency) and $\hbar\omega_{\rm L}=2.33~\textrm{eV}$ is the laser photon energy.

For sample 2 (see Fig.~\ref{Fig7} and Table~\ref{table1}), a value of $\lambda_\Gamma=0.034$  is deduced from the simultaneous fits of $\Delta\omega_{\rm G}$ and $\Delta\Gamma_{\rm G}$  (see Sec.~\ref{sec5}). Then, using Eq.~\eqref{eq_scattering}, we can estimate\footnote{Following Refs.~\onlinecite{Basko2009,Basko2008b}, the value of $\lambda_{K}$ deduced from Eq.~\eqref{eq_scattering} corresponds to the electron-phonon coupling constant at a carrier energy of $\hbar\omega_{\rm L}/2$.  To obtain the coupling constant exactly at the K point $\lambda_\textrm{K}(\hbar\omega_{\rm K})$, we can use the relation $\lambda_\textrm{K}(\hbar\omega_\textrm{K})/\lambda_\textrm{K}(\hbar\omega_{\rm{L}}/2)\approx 1.2$ that is valid for a polymer electrolyte with a relative permittivity $\varepsilon_R\approx 5$. In this manuscript, $\lambda_\textrm{K}$ implicitly denotes $\lambda_\textrm{K}(\hbar\omega_{\rm L}/2)$.} $\lambda_\textrm{K}= 0.17$. Overall, for the three pristine samples studied here, we obtained average values of $\left<\lambda_{\Gamma}\right>=0.031~\pm~0.004 $, $\left<\lambda_{\rm K}\right>=0.15~\pm~0.03$ and $\left<\frac{\lambda_{\rm K}}{\lambda_{\Gamma}}\right> = 5.1~\pm~1.2$  (see Table~\ref{table1}). 

To close this section, we compare the average ratio $\left<\frac{\lambda_\textrm{K}}{\lambda_{\Gamma}}\right>$ deduced from our doping-dependent Raman study to a direct estimate derived from the measured ratio of the integrated intensities of the intravalley (2D' mode) and intervalley (2D mode) resonant two-phonon features.\cite{Ferrari2013} This ratio is expected to be independent of $E_\textrm{F}$ and writes\cite{Basko2008,Basko2009} $\frac{I_\textrm{2D}}{I_\textrm{2D'}}=2\left(\frac{\lambda_{\rm K}}{\lambda_\Gamma}\right)^2$. 
In our experimental conditions, we obtain $\frac{I_\textrm{2D}}{I_\textrm{2D'}}=40~\pm~2$. Thus, by considering one more time the different Raman enhancement factors for the 2D- and 2D'-mode features in the PEO/graphene/SiO$_2~(285~\rm nm)$/Si multilayer system, we deduce $\frac{\lambda_{\rm K}}{\lambda_{\Gamma}}\approx 4.2$. This value is consistent with the analysis outlined above.

\begin{table*} [ht!]
\begin{center}
\begin{tabular}{ccccccc}
\hline
\hline
~~\textbf{Sample}~~ & ~~ $\boldsymbol{\left.\frac{I_{\rm 2D}}{I_{\rm G}}\right|_\textrm{intr}}$ ~~ & ~~ $\boldsymbol{\left(\frac{I_{\rm D}}{I_{\rm G}}\right)_0}$~~ & ~~ $\boldsymbol{n_\textrm{D}(\times 10^{12}\textrm{cm}^{-2})}$ ~~ & ~~$\boldsymbol{\gamma_\textrm{\textbf{e-ph}} + \gamma_\textrm{\textbf{D}}~(\textrm{\textbf{meV}})}$~~ & ~~ $\boldsymbol{\lambda_\Gamma}$ ~~ & ~~ $\boldsymbol{\lambda_\textrm{\textbf{K}}}$~~   \\
\hline
\hline
1 (without defects) &  5.6 & $<0.05$ & - & 50 & 0.027 &~ 0.17   \\

1 (with defects) & 4.6 & 1.7 & 0.9 & 57 & 0.031 & $~\lesssim 0.20$   \\

2 (without defects)  & 4.2 & $<0.05$ & - & 51 & 0.034 & ~0.17  \\

2 (with defects) & 3.3  & 1.3 & 0.7 & 69 & 0.031 & $~\lesssim 0.24$ \\

3 & 5.0 & $<0.05$ & - & 39 & 0.031 &~ 0.12   \\

4 & 4.4 & 2.6 & 1.4 & 53 & 0.037 & $~\lesssim 0.18$   \\

5 & 4.3 & 1.4 & 0.7 & 72 & 0.031 & $~\lesssim 0.25$   \\
\hline
\hline
\end{tabular}
\end{center}  
\caption{\textit{Intrinsic} integrated intensity ratio (corrected from interference effects~\cite{Yoon2009}) $\left.\frac{I_{\rm 2D}}{I_{\rm G}}\right|_{\rm intr}$ at $E_{\rm F}=0$, measured integrated intensity ratio $\left(\frac{I_{\rm D}}{I_{\rm G}}\right)_0$ at $E_{\rm F}=0$, estimated defect concentration $n_{\rm D}$, sum of the electron-phonon and electron-defect scattering rates, and dimensionless electron-phonon coupling constants at $\Gamma$ and near K (K'), for five different electrochemically gated graphene transistors. The measurements before and after the creation of defects have been done (i) at the same spot in sample 1 and correspond to the data shown in Fig.~\ref{Fig10}, and (ii) at two different spots in sample 2.}
\label{table1}
\end{table*}

\section{Defective graphene}
\label{sec8}

\subsection{Creation of defects}
\label{sec8A}

\begin{figure}[!tb]
\begin{center}
\includegraphics[width=8.6cm]{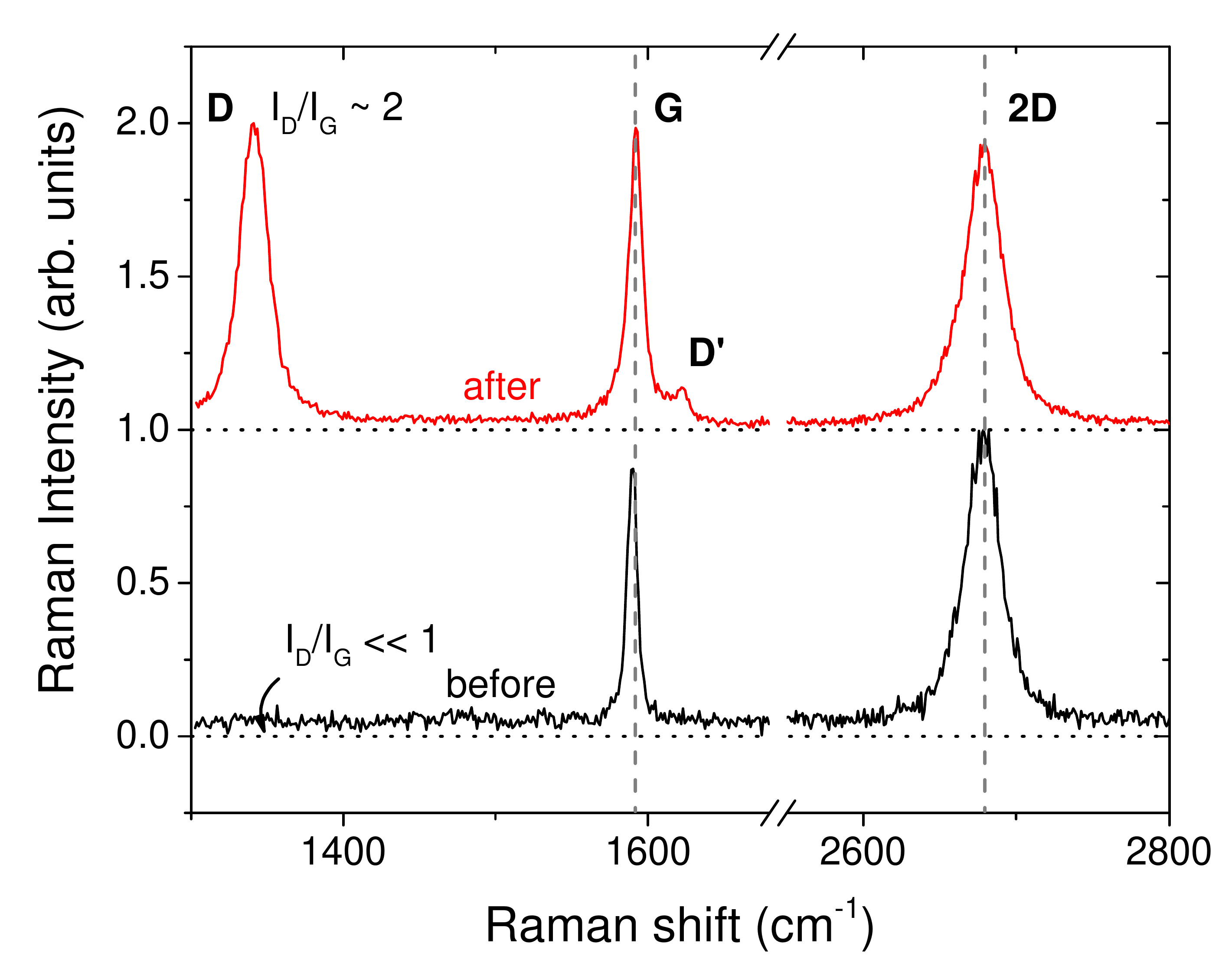}
\caption{Raman spectra at $V_{\rm TG}=0$ recorded at the same location on sample 1, before any gate bias has been applied (black line) and after an electrochemical reaction has occurred (red line). The spectra are vertically offset for clarity. The dotted lines correspond to the baseline.}
\label{Fig8}
\end{center}
\end{figure}

As mentioned in Sec.~\ref{sec2}, when an electrochemically gated graphene FET is subjected to a sufficiently high gate bias, electrochemical reactions may occur~\cite{Kalbac2010,Efetov2010,Bruna2014} and create defects in the graphene channel. In our devices, a reaction systematically occurs at negative gate biases ($V_{\rm TG}\approx-1~\textrm{V}$ to $-2~\textrm{V}$). The threshold voltage depends on the sample and on the gate capacitance. Electrochemical reactions result in an increase of the gate leak current above $1~\textrm{nA}$, and in the emergence of defect-induced features in the Raman spectrum. Figure \ref{Fig8} shows two Raman spectra recorded at $V_{\rm TG}=0~\textrm{V}$ on sample 1, before applying any gate voltage and after an electrochemical reaction has taken place. We clearly see that (i) the G- and 2D-mode features do not shift, and (ii) prominent D- and D'-mode features develop. These two Raman modes are known to be forbidden by symmetry and can only be observed in the presence of defects. \cite{Ferrari2000,Thomsen2000,Maultzsch2004,Malard2009,Ferrari2013}

\subsection{Doping dependence of the Raman features in defective graphene}
\label{sec8B}

\begin{figure}[!tb]
\begin{center}
\includegraphics[width=8.6cm]{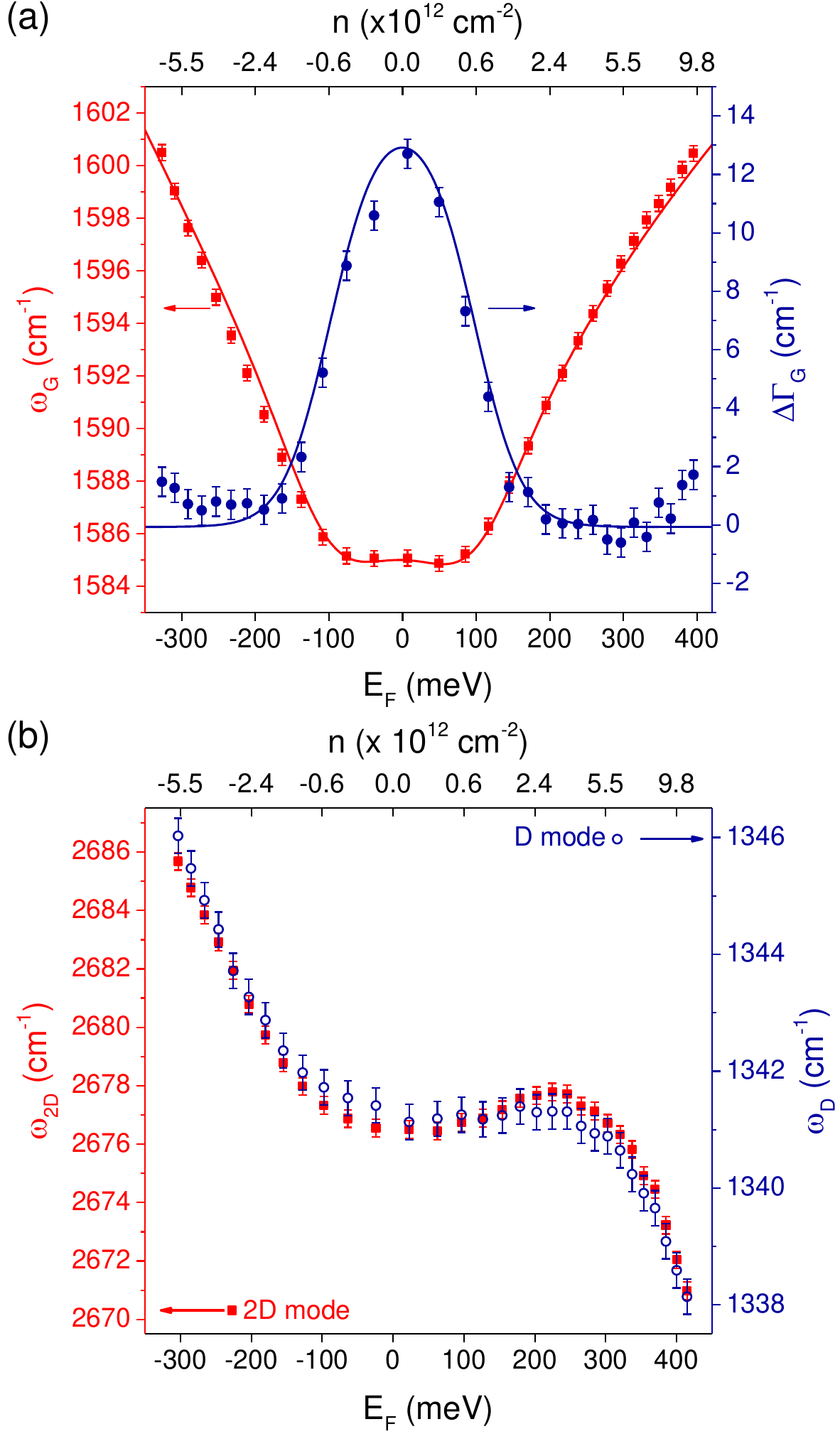}
\caption{Doping dependence of the Raman features in sample 1, after the creation of defects. (a) Doping dependence of the frequency $\omega_\textrm{G}$ (red squares, left axis) and relative FWHM $\Delta\omega_\textrm{G}$ (blue circles, right axis) of the G-mode feature. The blue and red solid lines are fits based on Eq.~\eqref{eq_fwhm} and \eqref{eq_posG}, as in Fig.~\ref{Fig4}, respectively. The fitting parameters are $C_\textrm{TG}=3.4~\mu\textrm{F~cm}^{-2}$, $\lambda_\Gamma=0.035$ and $\delta E_\textrm{F} = 30~\textrm{meV}$ (b) Doping dependence of the frequencies of the 2D- (red squares) and D-mode (blue open circles) features (denoted $\omega_\textrm{2D}$ and $\omega_\textrm{D}$, respectively) in defective graphene.}
\label{Fig9}
\end{center}
\end{figure}

Figure \ref{Fig9}(a) shows $\omega_{\rm G}(E_{\rm F})$ and $\Delta\Gamma_{\rm G}(E_{\rm F})$ in defective graphene. By comparing this figure with Fig.~\ref{Fig4}, we conclude that the doping dependence of the G-mode feature is not affected by the presence of defects. Both $\Delta\omega_{\rm G}(E_{\rm F})$ and $\Delta\Gamma_{\rm G}(E_{\rm F})$ are well fit to the theoretical model of Sec.~\ref{sec5}. The frequencies $\omega_{\rm 2D}$ and $\omega_{\rm D}$ are also shown as a function of $E_{\rm F}$ in Fig.~\ref{Fig9}(b). The D- and 2D-mode features are fit to a modified Lorentzian profile.\cite{Basko2008,Berciaud2013} We note that both frequencies follow identical trends. More precisely $\omega_{\rm 2D}\approx 2\omega_{\rm D}-6~\textrm{in cm}^{-1}$. The factor 2 is expected, since the 2D mode is the two-phonon overtone of the D-mode (in the D-mode process, one inelastic scattering by a near zone-edge TO phonon is replaced by an elastic scattering by a defect~\cite{Thomsen2000,Basko2008,Venezuela2011}). This small difference between $2\omega_{\rm D}$ and $\omega_{\rm 2D}$ is consistently observed in all the studied samples. It could be due to slight differences in the resonance conditions.\cite{MartinsFerreira2010,Venezuela2011,Ferrari2013}

\subsection{Electron-defect scattering}
\label{sec8C}

\begin{figure*}[!hbt]
\begin{center}
\includegraphics[width=17.8cm]{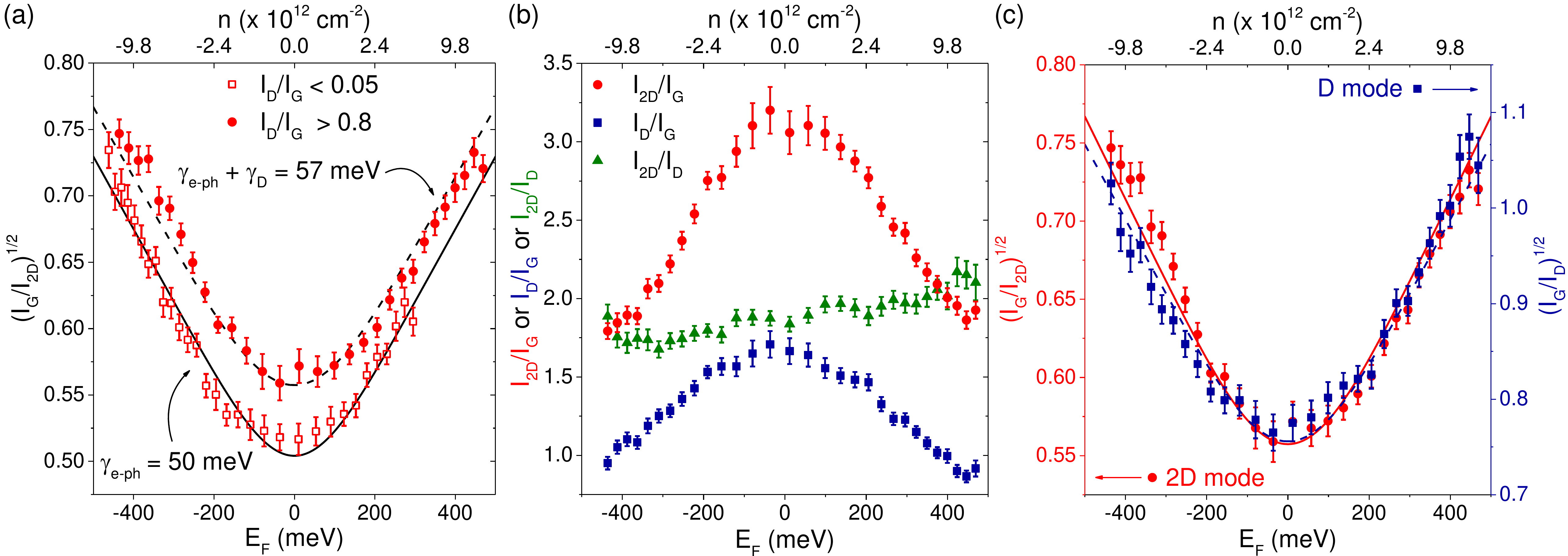}
\caption{(a) Integrated intensity ratio $\sqrt{\frac{I_{\rm G}}{I_{\rm 2D}}}$ as a function of $E_{\rm F}$ in sample 1 before (red open squares) and after (red circles) the creation of defects. (b) Integrated intensity ratios $\frac{I_{\rm 2D}}{I_{\rm G}}$ (red circles), $\frac{I_{\rm 2D}}{I_{\rm D}}$ (green triangles) and $\frac{I_{\rm D}}{I_{\rm G}}$ (blue squares), as a function of $E_{\rm F}$ in sample 1, after the creation of defects. (c)  $\sqrt{\frac{I_{\rm G}}{I_{\rm 2D}}}$ (red circles, left axis) and $\sqrt{\frac{I_{\rm G}}{I_{\rm D}}}$ (blue squares, right axis) as a function of $E_{\rm F}$, in sample 1, after the creation of defects. As described in Sec.~\ref{sec8D}, a defect concentration $n_{\rm D}\approx9\times 10^{11}~\textrm{cm}^{-2}$ is estimated from these measurements.}
\label{Fig10}
\end{center}
\end{figure*}

In Sec.~\ref{sec7}, using Eq.~\eqref{eq_ratio}, we have shown that it is possible to deduce the phonon scattering rate $\gamma_{\textrm{e-ph}}$ in pristine graphene from the study of the integrated intensities of the G- and 2D-mode features. In defective graphene, a similar analysis can be performed provided that a finite electron-defect scattering rate $\gamma_{\rm D}$, proportional to the defect concentration $n_{\rm D}$, is taken into consideration.\cite{Venezuela2011,Bruna2014}

In Fig.~\ref{Fig10}(a), we plot $\sqrt{\frac{I_{\rm G}}{I_{\rm 2D}}}$ as a function of $E_{\rm F}$. The data is extracted from another series of measurements in sample 1, at a same spot, before and after the creation of defects. As expected, we observe that the two datasets are well fitted by Eq.~\eqref{eq_ratio}. The fitting parameters are $\gamma_{\textrm{e-ph}}= 50~\textrm{meV}$, $\left.\frac{I_{\rm 2D}}{I_{\rm G}}\right|_0^{\textrm{no D}} =4.8$, $\delta E_{\rm F} = 110~\textrm{meV}$ for pristine graphene, and $\gamma_{\textrm{e-ph}}+\gamma_{\rm D} = 57~\textrm{meV}$, $\left.\frac{I_{\rm 2D}}{I_{\rm G}}\right|_0^{\textrm{D}}  = 4.0$, $\delta E_{\rm F} = 130~\textrm{meV}$, for defective graphene, respectively. 

Since $\gamma_{\textrm{e-ph}}$ is not affected by the presence of defects, we can estimate that $\gamma_{\rm D}\approx 7~\textrm{meV}$ for sample 1. Thus, although the D- and 2D-mode features have similar integrated intensities ($I_{\rm D}/I_{\rm 2D}\approx 0.5$, as shown in Fig.~\ref{Fig10}(b)), $\gamma_{\rm D}\ll\gamma_{\textrm{e-ph}}$, as predicted in the low-defect concentration regime \cite{Venezuela2011} (see also Sec.~\ref{sec8D}). Another way to determine $\gamma_{\textrm{D}}$ is to compare the quantity $\left.\sqrt{\frac{I_{\rm G}}{I_{\rm 2D}}}\right|_0$, in the presence and in the absence of defects. According to Eq.~\eqref{eqI2D} and \eqref{eq_ratio}, one obtains
\begin{equation}
\frac{\left.\sqrt{\frac{I_{\rm G}}{I_{\rm 2D}}}\right|_0^{\textrm{D}}}{\left.\sqrt{\frac{I_{\rm G}}{I_{\rm 2D}}}\right|_0^{\textrm{no D}}}=1+\frac{\gamma_{\rm D}}{\gamma_{\textrm{e-ph}}}.
\label{eq18}
\end{equation}
The results of Fig.~\ref{Fig10}(a) indeed show that $\left.\sqrt{\frac{I_{\rm G}}{I_{\rm 2D}}}\right|_0^{\textrm{D}}~>~\left.\sqrt{\frac{I_{\rm G}}{I_{\rm 2D}}}\right|_0^{\textrm{no D}} $, in agreement with Eq.~\eqref{eq18}. From the fitting parameters, we estimate that $\gamma_{\rm D}\approx 5 ~\textrm{meV}$, in good agreement with the other estimate obtained above.

To conclude this subsection, we focus on the dependence of $I_{\rm D}$ on $E_{\rm F}$. In practice, $I_{\rm D}$ is routinely used to estimate a defect concentration.\cite{Tuinstra1970,Ferrari2000,MartinsFerreira2010,Lucchese2010,Cancado2011,Eckmann2013,Ferrari2013,Beams2015} However, although it is not \textit{fully resonant}, the D mode may involve one resonant electron-phonon scattering process~\cite{Reich2000,Maultzsch2004,Ferrari2013,Venezuela2011,Beams2015} (see Fig.~\ref{Fig2}(g)). In other words, similarly to $I_{\rm 2D}$, $I_{\rm D}$ is also expected to decrease with increasing $\abs{E_{\rm F}}$. In Fig.~\ref{Fig10}(b), we show $\frac{I_{\rm 2D}}{I_{\rm G}}$, $\frac{I_{\rm 2D}}{I_{\rm D}}$ and $\frac{I_{\rm D}}{I_{\rm G}}$ as a function of $E_{\rm F}$, while Fig.~\ref{Fig10}(c) displays $\sqrt{\frac{I_{\rm G}}{I_{\rm 2D}}}$ and $\sqrt{\frac{I_{\rm G}}{I_{\rm D}}}$ as a function of $E_{\rm F}$. Clearly, $I_{\rm D}$ and $I_{\rm 2D}$ show a very similar doping dependence (see also Ref.~\onlinecite{Bruna2014}). More quantitatively, a phenomenological fit of $\sqrt{\frac{I_{\rm G}}{I_{\rm D}}}(E_{\rm F})$ using Eq.~\eqref{eq_ratio} (applied to $I_{\rm D}$ instead of $I_{\rm 2D}$) agrees well with our measurements (see Fig.~\ref{Fig10}(c)) and yields $\gamma_{\textrm{e-ph}}+\gamma_{\rm D} = 52~\textrm{meV}$, $\left.\frac{I_{\rm D}}{I_{\rm G}}\right|_0 = 2.2 $, and $\delta E_{\rm F} = 130~\textrm{meV}$. The value of $\gamma_{\textrm{e-ph}}+\gamma_{\rm D}$ is very close to that obtained by fitting $\sqrt{\frac{I_{\rm G}}{I_{\rm 2D}}}(E_{\rm F})$.

We note that Ref.~\onlinecite{Bruna2014} report a value of $\gamma_{\rm e-ph}+\gamma_{\rm D}\sim 70~\rm meV$ similar to ours (see Table~\ref{table1}), for defective graphene samples with slightly larger, yet similar values of $\left.\frac{I_{\rm D}}{I_{\rm G}}\right|_0$. However, a larger value of $\gamma_{\rm D}\sim 40~\rm meV$ is estimated, using the value of $\gamma_{\rm{e-ph}}\sim 30~\rm meV$ extracted from the measurements in Ref.~\onlinecite{Basko2009}. Our work provides an estimate of $\gamma_{\rm D}$ from a series of measurements performed on a \textit{same} sample and suggests that $\gamma_{\rm D}\ll \gamma_{\rm e-ph}$, even for $\left.\frac{I_{\rm D}}{I_{\rm G}}\right|_0\gtrsim 1$. Consequently, in defective graphene, we consider that the slope of the  $\sqrt{\frac{I_{\rm G}}{I_{\rm 2D}}}(E_{\rm F})$ curve provides a fair estimate of $\gamma_{\rm e-ph}$, from which we extract $\lambda_{\rm K}$, knowing $\lambda_{\Gamma}$. Albeit the existence of a finite $\gamma_{\rm D}$ presumably leads to a slight overestimation of $\lambda_{\rm K}$, we do not observe a large difference between the values measured on defective and on pristine graphene (see Table~\ref{table1}). More quantitatively, by averaging on four defective graphene samples with similar defect concentrations, we obtain $\left <\gamma_{\rm e-ph}+\gamma_{\rm D} \right>=(63~\pm~9)~\rm meV$, a value that is indeed slightly larger than the average $\left <\gamma_{\rm e-ph} \right>=(47~\pm~7)~\rm meV$ obtained on three pristine samples (see Table~\ref{table1} and Sec.~\ref{sec7}).

\subsection{Defect concentration}
\label{sec8D}

In principle, the concentration $n_{\rm D}$ of defects in a graphene sample can be deduced from the analysis of the defect-induced Raman modes, such as the (intervalley) D mode or the (intravalley) D' mode. The study of the defect-induced Raman modes has far reaching consequences for sample characterization and can also be a very useful tool to monitor chemical reactions on graphene.  Following the seminal work by Tuinstra and Koenig,\cite{Tuinstra1970} several groups have proposed analytical expressions to connect $I_{\rm D}$ and $n_{\rm D}$ in various graphitic materials, from weakly defective graphene layers to amorphous carbon.\cite{Tuinstra1970,Ferrari2000,MartinsFerreira2010,Lucchese2010,Cancado2011}

\begin{figure}[!tb]
\begin{center}
\includegraphics[width=8.6cm]{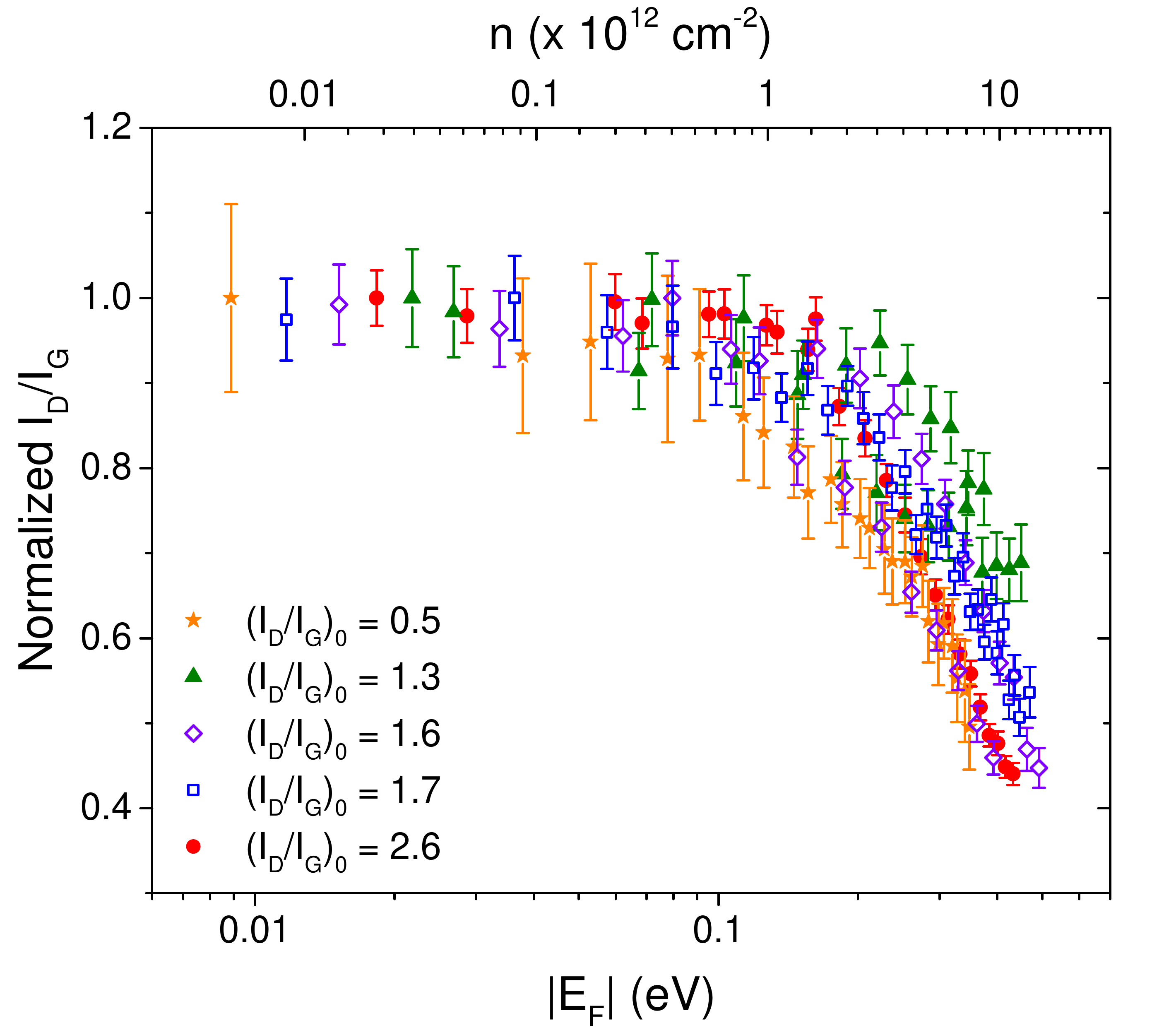}
\caption{Doping dependence of the integrated intensity ratio $\frac{I_{\rm D}}{I_{\rm G}}$, for five sets of measurements on defective graphene samples. Each dataset is normalized by the value $\left(\frac{I_{\rm D}}{I_{\rm G}}\right)_0$ measured near the charge neutrality point. The measured $\left(\frac{I_{\rm D}}{I_{\rm G}}\right)_0$ are indicated in the legend.}
\label{Fig11}
\end{center}
\end{figure}

The defective graphene samples studied here exhibit an integrated intensity ratio $\frac{I_{\rm D}}{I_{\rm G}}\leq2.6$ near the CNP (see Fig.~\ref{Fig8} and \ref{Fig11}, and Table~\ref{table1}). Their Raman features show only a slight spectral broadening (by a few $\textrm{cm}^{-1}$) compared to the pristine case (see Fig.~\ref{Fig8}). More precisely, we find, for five series of gate-dependent measurements on various defective regions, that $\Gamma_{0}\sim 10~\textrm{cm}^{-1}$ (as compared to $\Gamma_{0}\approx 5~\textrm{cm}^{-1}$ for pristine graphene), $\Gamma_{\rm 2D}\sim 30~\textrm{cm}^{-1}$ and $\Gamma_{\rm D}\sim 20~\textrm{cm}^{-1}$.  Following the three stage classification of Ref.~\onlinecite{Ferrari2000} and related works,\cite{MartinsFerreira2010,Lucchese2010,Cancado2011,Eckmann2013} such samples can be described as stage 1, \textit{i.e.,} still in the weakly defective regime. Let us assume point defects, separated by an average distance $L_{\rm D}\gtrsim10~\textrm{nm}$. In this regime, Eq.~(9) of Ref.~\onlinecite{MartinsFerreira2010} and the results of Ref.~\onlinecite{Cancado2011} provide the relation\footnote{We note that since the D- and G-mode frequencies are relatively close, the impact of interference effects on the $\frac{I_{\rm D}}{I_{\rm G}}$ ratio is negligible in our experimental conditions.\cite{Yoon2009}}

\begin{equation}
n_{\rm D}=\frac{10^{14}}{L_{\rm D}^2}\approx1.8\times 10^{10} (\hbar\omega_{\rm L})^4 \left(\frac{I_{\rm D}}{I_{\rm G}}\right)_0,
\label{eq_nd}
\end{equation}
where $n_{\rm D}$ is the concentration of defects in cm$^{-2}$, $L_{\rm D}$ is in nm, $\hbar\omega_{\rm L}$ is the laser photon energy in eV and $\left(\frac{I_{\rm D}}{I_{\rm G}}\right)_0$ is taken at $E_{\rm F}\approx 0$, still with Fermi energy fluctuations. According to Ref.~\onlinecite{Eckmann2013}, the scaling introduced in Eq.~\eqref{eq_nd} is independent of the type of defect.

In Fig.~\ref{Fig11}, we plot $\frac{I_{\rm D}}{I_{\rm G}}\left(E_{\rm F}\right)$, normalized by its value at  $E_{\rm F}\approx 0$ for five different sets of measurements (including some at different locations on the same sample), with different defect concentrations. We observe that all the data collapse onto the same curve. For $\abs{E_{\rm F}}\lesssim100~\textrm{meV}$, $\frac{I_{\rm D}}{I_{\rm G}}\approx\left(\frac{I_{\rm D}}{I_{\rm G}}\right)_0$ and this ratio decreases by less than $20~\%$ for $\abs{E_{\rm F}}\lesssim 200~\textrm{meV}$. Thus, since unintentional doping in graphene samples typically leads to $\abs{E_{\rm F}}\lesssim 200~\textrm{meV}$, the experimentally measured $\frac{I_{\rm D}}{I_{\rm G}}$ can be used together with  Eq.~\eqref{eq_nd} for an estimation of $n_{\rm D}$ in weakly doped samples.

Using Eq.~\eqref{eq_nd}, the electrochemically-induced defect concentrations deduced for our measurements (see Table~\ref{table1} and Fig.~\ref{Fig11}) range from $2.7\times 10^{11}~\textrm{cm}^{-2}$ to $1.4\times 10^{12}~\textrm{cm}^{-2}$. This translates into $L_{\rm D}$ ranging from 19.5~nm down to 8.5~nm. The latter value is at the limit of the weakly defective regime, which assumes $L_{\rm D}\gtrsim10~\rm nm$.

Overall, the results shown in Figs.~\ref{Fig9}-\ref{Fig11} demonstrate that for defect concentrations below approximately $ 2\times10^{12}~\rm cm^2$, the electron-defect scattering rate remains much smaller than the electron-phonon scattering rate, and that the doping dependence of the G- and 2D-mode features is essentially the same as in pristine graphene. These results contrast with the fact that even for relatively low $n_{\rm D}$ in the range $10^{11}-10^{12}~\rm cm^{-2}$, the integrated intensity of the D-mode feature is smaller, yet on the same order of magnitude as that of the 2D-mode feature, in keeping with recent experimental\cite{MartinsFerreira2010,Bruna2014} and theoretical results.\cite{Venezuela2011} This calls for further investigations of the integrated intensity of the one-phonon, defect-induced Raman features relative to that of their symmetry-allowed overtones.

\section{Correlations}
\label{sec6}

In the previous sections, we have successfully compared our measurements to theoretical calculations and, in particular estimated the electron-phonon coupling constants. In this section, we present correlations between the frequencies and linewidths of the main Raman features in doped graphene, with the aim to extract universal behaviors that could be useful for sample characterization. Based on the conclusions of Sec.~\ref{sec8}, the correlations discussed in the following will also hold in weakly defective graphene.

\subsection{G-mode frequency and linewidth}
\label{sec6A}

\begin{figure}[!tb]
\begin{center}
\includegraphics[width=8.6cm]{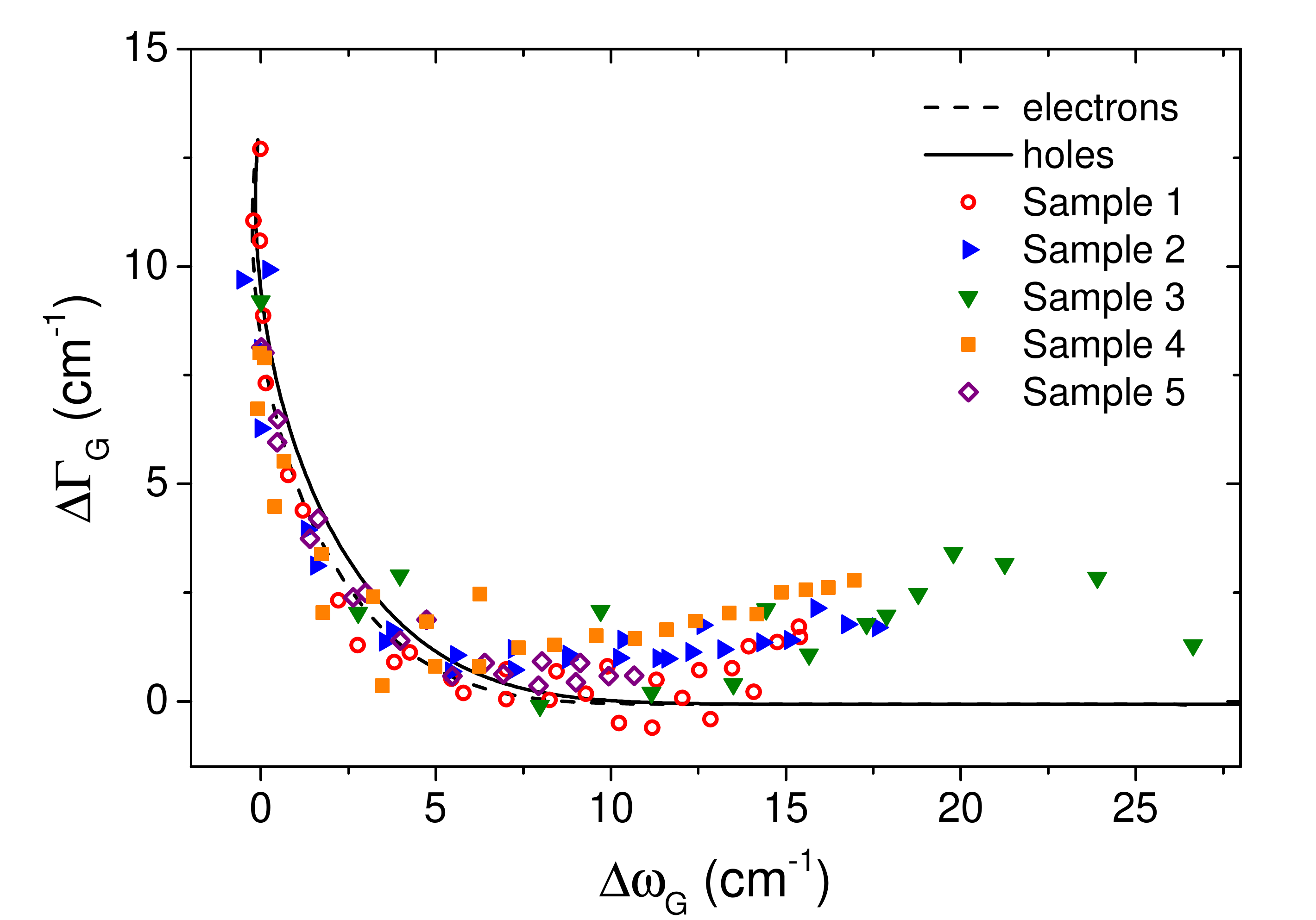}
\caption{Correlation between the relative FWHM $\Delta\Gamma_{\rm G}$ and relative frequency $\Delta\omega_{\rm G}$ of the G-mode feature in doped graphene, for the five samples introduced in Fig.~\ref{Fig5}. The solid and dashed lines correspond to theoretical calculations (based on Eqs.~\eqref{eq_fwhm} and \eqref{eq_posG}), for electron and hole doping, respectively.}
\label{Fig12}
\end{center}
\end{figure}

Figure~\ref{Fig12} shows $\Delta\Gamma_{\rm G}$ as a function of $\Delta\omega_{\rm G}$ for the five different samples already shown in Fig.~\ref{Fig5}. We observe a universal behavior and the experimental data are in good agreement with the theoretical calculations, although the very slight difference expected for electron and hole doping (due to $\Delta\omega_{\rm G}^{\rm A}$, see Eq.~\eqref{eq_posG}) is not resolved experimentally, likely due to Fermi energy fluctuations. We also note that in the high-doping regime, $\Delta\Gamma_{\rm G}$ tends to increase somewhat. This increase, also observed by others,\cite{Bruna2014} is presumably due to the increasing inhomogeneity of the charge distribution at high top-gate biases. The correlation displayed in Fig.~\ref{Fig12}  may also be used to estimate $E_{\rm F}$, especially in the low doping regime $\left(\abs{E_{\rm F}}\lesssim\hbar\omega_{\rm G}^0\right)$.

\subsection{G- and 2D-mode frequencies}
\label{sec6B}

\begin{figure}[!tb]
\begin{center}
\includegraphics[width=8.6cm]{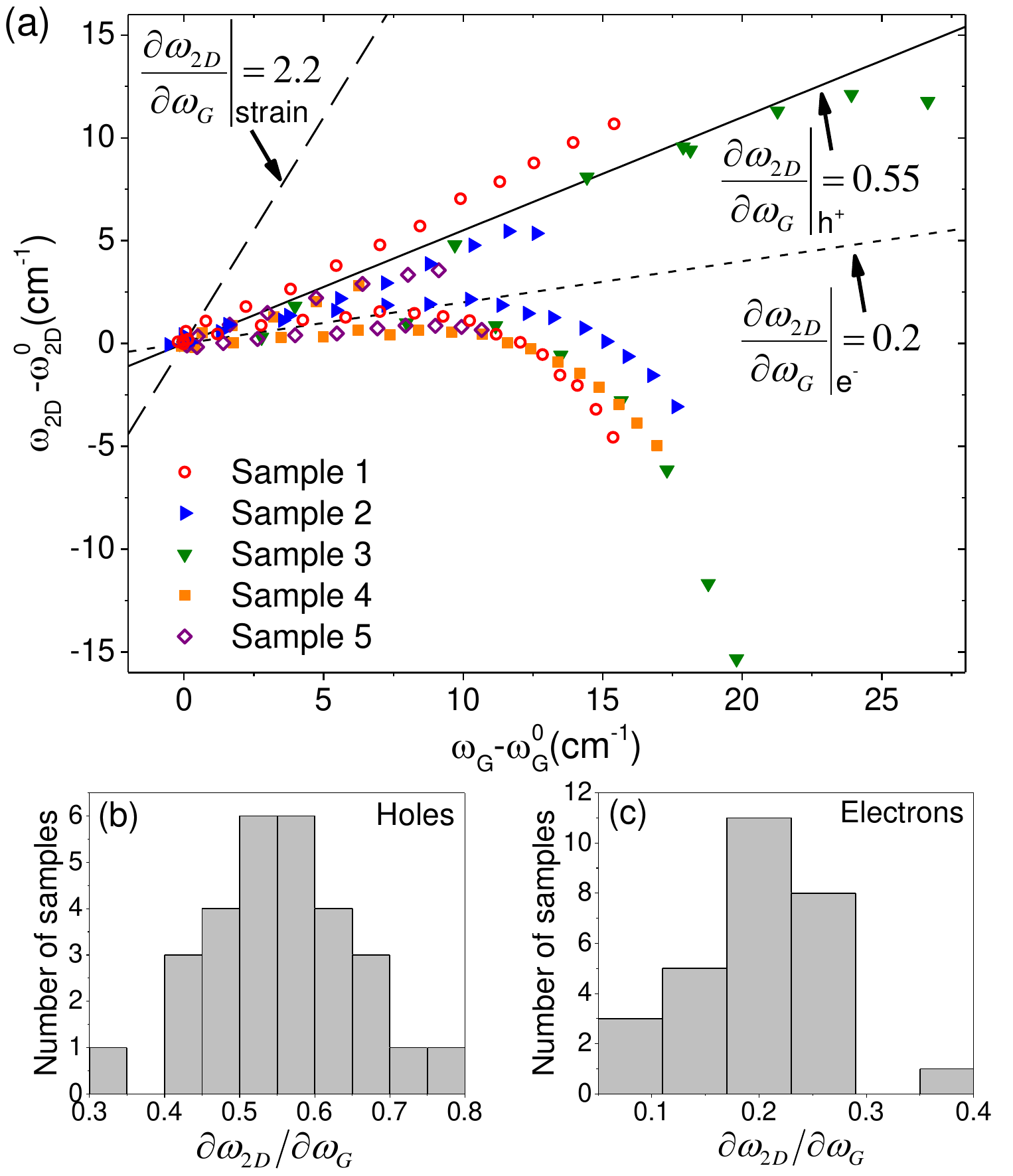}
\caption{(a) Correlation between the frequencies of the 2D- and G-mode features (relative to their values near the charge neutrality point) in doped graphene, for the five samples introduced in Fig.~\ref{Fig5}. The continuous and short-dashed lines are global fits performed on the linear portions of the hole doping and electron doping branches, respectively. The dashed line corresponds to the evolution of  $\omega_{\rm 2D}$ versus $\omega_{\rm G}$ under pure strain.\cite{Lee2012,Metten2014} Statistical distribution of the measured slopes $\frac{\partial\omega_{\rm 2D}}{\partial \omega_{\rm G}}$ under hole and electron doping are shown in (b) and (c), respectively.}
\label{Fig13}
\end{center}
\end{figure}

Figure \ref{Fig13} represents the evolution of $\omega_{\rm 2D}$ as a function of $\omega_{\rm G}$ for the same five samples. A clear correlation is observed between these two quantities. For hole doping, the correlation is quasi-linear in the range of $E_{\rm F}$ studied here $(-500~\textrm{meV}\lesssim E_\textrm{F}<0)$. In contrast, for electron doping, a quasi-linear scaling, again with a (much smaller) positive slope is also observed at low doping $(0<E_{\rm F}\lesssim 250~\textrm{meV})$, until $\omega_{\rm 2D}$ levels off and ultimately decreases, leading to a non-linear scaling. This behavior was observed on every sample either for electrolyte-gated or conventional back-gated FETs and has been also observed in chemically doped graphene.\cite{Lee2012}

From the slopes $\frac{\partial\omega_{\rm 2D}}{\partial\omega_{\rm G}}$ extracted on approximately thirty samples (see Fig.~\ref{Fig13}(b)-(c)), we find an average of $(0.55~\pm~0.2)$ for hole doping and of $(0.2~\pm~0.13)$ for electron doping, respectively. The former value agrees well with the slope of $(0.70~\pm~0.05)$ extracted numerically by Lee \textit{et al.}\cite{Lee2012} from the data in Ref.~\onlinecite{Das2008,Das2009}. From our statistical study, we note that  the correlation between $\omega_{\rm 2D}$ and $\omega_{\rm G}$ is more dispersed than the correlation between $\Delta\Gamma_{\rm G}$ and $\Delta\omega_{\rm G}$. This is chiefly due to the dependence of $\omega_{\rm 2D}$ on $E_{\rm F}$, which is not as universal as that of $\omega_{\rm G}$. In addition, it is rather challenging to extract a well-defined correlation for electron doping due to the small variations of $\omega_{\rm 2D}$ at moderate doping.

 Noteworthy, estimations of $E_{\rm F}$ based on the frequency and/or linewidth of the Raman features may only be reliable if graphene is not subjected to significant strains. Indeed, $\Gamma_{\rm G}$ is marginally affected by isotropic strains below $1\%$.\cite{Metten2014} However, the G-mode feature may broaden and ultimately split into two sub-features in the presence of larger anisotropic strains.\cite{Mohiuddin2009,Huang2009} In addition, the Raman features soften (stiffen) under tensile (compressive) strain. A linear correlation between $\omega_{\rm 2D}$ and $\omega_{\rm G}$ has been measured in strained graphene.\cite{Lee2012,Zabel2012,Lee2012NL,Metten2013,Metten2014} Since the measured slopes ($\frac{\partial\omega_{\rm 2D}}{\partial\omega_{\rm G}}\approx2.2$, for undoped, strained graphene~\cite{Metten2014}) are appreciably larger than the slopes measured in doped graphene (presumably under a small but constant built-in strain), Lee \textit{et al.} have proposed to use the correlation between $\omega_{\rm 2D}$ and $\omega_{\rm G}$ as a robust tool to optically separate strain from charge doping.\cite{Lee2012}

\begin{figure}[!tb]
\begin{center}
\includegraphics[width=8.6cm]{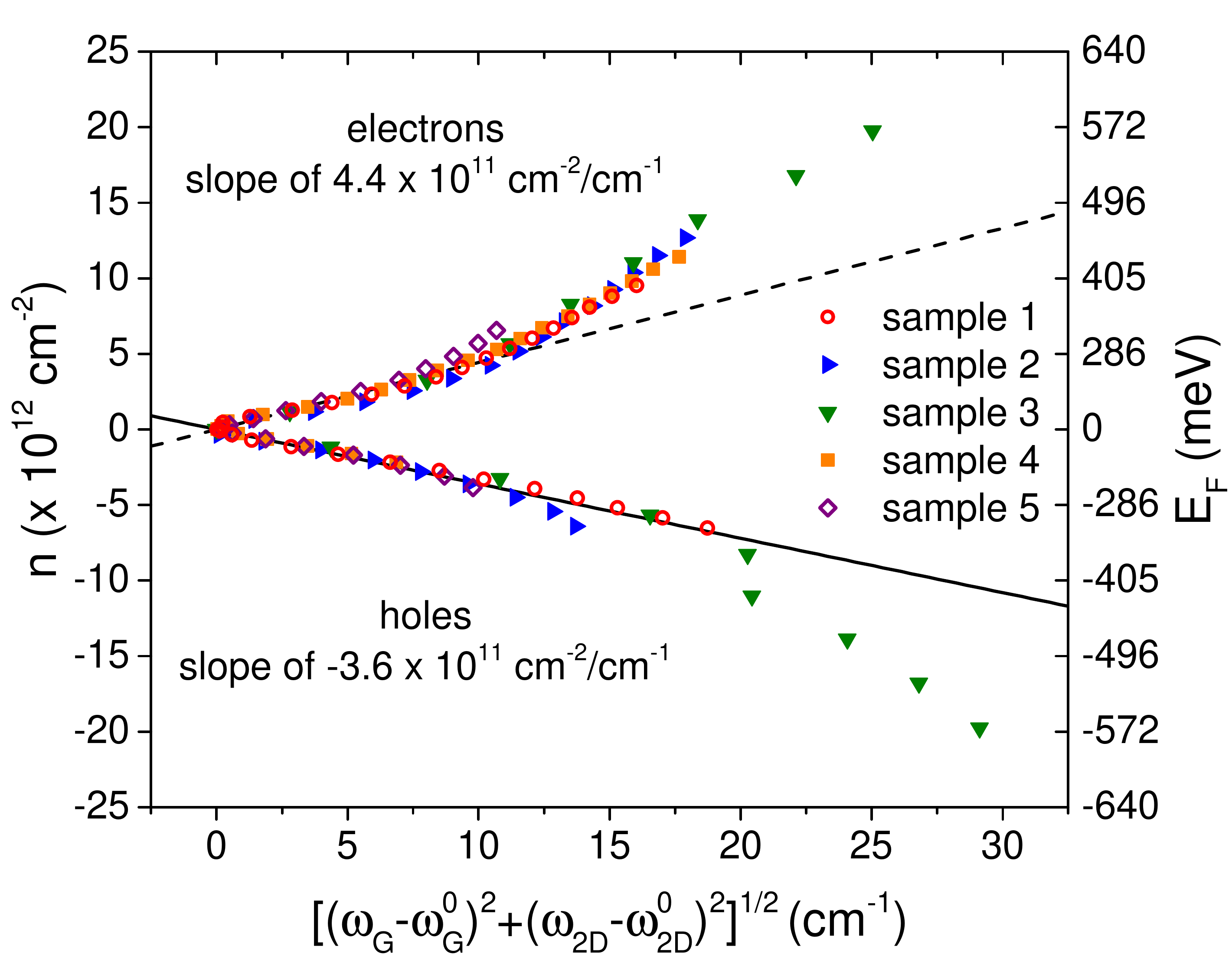}
\caption{Charge carrier density $n$ as a function of the shift $\left[(\omega_{\rm G}-\omega_{\rm G}^0)^2+(\omega_{\rm 2D}-\omega_{\rm 2D}^0)^2\right]^{1/2}$ from the reference point corresponding to undoped graphene. Data are shown for the five  samples introduced in Fig.~\ref{Fig5}. The continuous and dashed lines are linear global fits performed on the electron and hole branches, respectively.}
\label{Fig14}
\end{center}
\end{figure}

Following Ref.~\onlinecite{Lee2012}, we may then define three vectors corresponding to the slopes $\frac{\partial\omega_{\rm 2D}}{\partial\omega_{\rm G}}$, under strain, hole and electron doping, respectively (see Fig.~\ref{Fig13}). To further deduce absolute levels of strain and/or doping, one also has to know the 2D- and G-mode frequencies that corresponds to an undoped and unstrained graphene sample. For clarity, in Fig.~\ref{Fig13},  the 2D- and G-mode frequencies are shown relative to the measurements at $E_{\rm F}\approx 0$. These origin points, denoted ($\omega_{\rm G}^0$, $\omega_{\rm 2D}^0$) might differ from the reference point corresponding to  undoped \textit{and} unstrained graphene, since an undetermined amount of native strain may be present and induce a shift along the strain vector.

The data in Fig.~\ref{Fig13} allows an estimation of the coefficient which connects $\left[(\omega_{\rm G}-\omega_{\rm G}^0)^2+(\omega_{\rm 2D}-\omega_{\rm 2D}^0)^2\right]^{1/2}$,  the measured distance from the \textit{zero doping} point, to a given doping level (see Fig.~\ref{Fig14}). We chose to consider $n$ instead of $E_{\rm F}$ because it is a more relevant quantity as far as graphene characterization is concerned. Although the curves displayed in Fig.~\ref{Fig14} are not expected to exhibit a linear scaling (as opposed to the data shown in Fig.~\ref{Fig5}), we observe a quasi-linear scaling for sufficiently small doping ($\abs{n}\lesssim7\times 10^{12}\rm cm^{-2}$). We therefore fit the linear part for both electron- and hole-doping with a line intercepting the \textit{zero doping} point. We find slopes of $4.4 \times 10^{11}~\textrm{cm}^{-1}$ for electrons and $-3.6 \times 10^{11}~\textrm{cm}^{-1}$ for holes, respectively.

Finally, considering the Gr\"uneisen parameters of 1.8 and 2.4 for the Raman G- and 2D-modes under isotropic strain,\cite{Metten2014} we can estimate a slope of $7.1 \times 10^{-3}~\%~\rm{strain}/\rm{cm^{-1}}$ to connect $\left[(\omega_{\rm G}-\omega_{\rm G}^0)^2+(\omega_{\rm 2D}-\omega_{\rm 2D}^0)^2\right]^{1/2}$ to an applied isotropic strain. In practice, the strain field may be anisotropic, depending on the sample and on the experimental conditions, leading to a different slope. These coefficients may be used for a reliable estimation of doping and strain in graphene samples and devices.

\section{Conclusion}

We have presented a robust method, based on Raman scattering spectroscopy, to accurately determine the geometrical capacitance, and hence, the Fermi energy in electrochemically-gated graphene field-effect transistors with a spatial resolution down to approximately $1~\mu \rm m$. Such a calibration allows for quantitative analysis of the doping dependence of the frequency, linewidth and integrated intensity of the main Raman features. The anomalous doping dependence of the G-mode phonons is well captured by theoretical models over a broad range of Fermi energies above or below the Dirac point, and provides an experimental measurement of the electron-phonon coupling constant at the $\Gamma$ point of the Brillouin zone. We have then exploited the peculiar doping dependence of the integrated intensity of the multiphonon resonant Raman features, in particular the resonant 2D-mode feature, to estimate the electron-phonon coupling constant at the edges (K, K') of the Brillouin zone. Finally, from the doping dependence of the integrated intensity of the defect-induced D-mode feature, we can estimate the electron-defect scattering rate in stage 1 defective graphene samples. 

Our study provides useful guidelines for the characterization of graphene samples. We have, in particular, considered the correlation between the frequency and width of the G-mode feature, as well as between the frequencies of the 2D- and G-mode features. These correlations reveal \textit{universal} behaviors that can therefore be applied to evaluate doping in a variety of experimental situations. We have also demonstrated that defects can be efficiently created \textit{in-situ} in electrochemically gated graphene field effect transistors. The integrated intensity of the D-mode feature decreases monotonically with increasing doping, and follows the same scaling as that of its two-phonon overtone. However, due to Fermi energy fluctuations, the D-mode intensity is nearly constant for Fermi energy shifts below 200~meV relative to the Dirac point, which is of practical interest for the determination of the defect concentration.

In the present work, we conservatively estimate that Fermi energies as high as $\approx 700~\rm meV$ above the Dirac point can be achieved in ambient conditions, without damaging graphene. This naturally opens exciting perspectives for optoelectronics. Nevertheless, a well controlled Raman scattering study of the crossover between the intermediate doping regime achieved here $(n\approx 10^{13}~\rm cm^{-2})$ and the very high doping regime $(n>10^{14}~\rm cm^{-2})$ remains elusive. Finally, electrochemical gating is a promising strategy to investigate electron-phonon coupling in other two-dimensional materials, including transition metal dichalcogenides.\cite{Chakraborty2012}

\begin{acknowledgments}
We wish to thank G. Weick for his careful proofreading of the manuscript and for fruitful discussions.
We are also grateful to D. M. Basko, F. Mauri, F. Federspiel, D. Metten, S. Zanettini and B. Doudin for discussions. We thank A. Mahmood, F. Godel, S. Kuppusamy, F. Chevrier, A. Boulard and M. Romeo for technical support, as well as R. Bernard,  S. Siegwald, and H. Majjad for help with sample preparation in the StNano clean room facility. We acknowledge support from the Agence Nationale de la Recherche (under grant QuanDoGra 12 JS10-001-01), from the CNRS and from Universit\'e de Strasbourg. 

\end{acknowledgments}


%

\end{document}